\documentclass{ccjnl}

\title{Joint Optimization of Resource Allocation and Radar Receiver Selection in Integrated Communication-Radar Systems}
\author{Chen Zhong\inst{1}, Xufeng Zhou\inst{1}, Lan Tang\inst{1,*},Mengting Lou\inst{2}\corinfo{tanglan@nju.edu.cn}}
\receiveddate{}
\reviseddate{}
\Editor{}

\address[1]{School of Electronic Science and Engineering, Nanjing University, Nanjing {\rm 210093}, China}
\address[2]{Future Research Lab, China Mobile Research Institute, Beijing {\rm 100053}, China}

\begin{document}
\maketitle

\begin{abstract}
Considering the advantages of multi-input multi-output and orthogonal frequency division multiplexing (MIMO-OFDM) in both communication and perception,  in this paper, we investigate  a distributed integrated MIMO-OFDM dual-function radar-communication (DFRC) system, in which communication and probing are implemented simultaneously in different subcarrier sets. We first design the beam pattern and transmission signals on sensing subcarriers, and derive Cramer-Rao Bound (CRB) of targets in detection area. Then, to obtain the best tradeoff between communication and sensing performance, we maximize the transmission rate by jointly optimizing the power/subcarrier allocation and the selection of radar receivers under the constraints of detection performance and total transmit power. To tackle the non-convex mixed integer programming problem, we propose an alternative optimization algorithm, which solves the original problem by solving two subproblems iteratively. For given radar receivers set, the first subproblem related to power/subcarriers allocation is transformed into a semidefinite programming (SDP) problem with difference of convex (DC) approximation. For given resource allocation, the second subproblem  related to radar receivers selection is converted to a convex quadratic integer problem. The numerical results demonstrate that tradeoff relationship between communication performance and radar performance as well as the performance improvement through radar receivers selection. 
\end{abstract}

\section{introduction}
\label{s1}

In recent years, to alleviate the congestion and shortage of radio frequency (RF) spectrum resource, wireless communication systems are gradually operating in the millimeter wave (mmWave) frequency band, which overlaps with the operating frequency band of the radar systems\cite{5776640}\cite{7019537}. To address the possible interference between communication and radar, some spectrum coexistence schemes of radar system and communication system have been researched\cite{8999605}. On the other hand, the similarity between radar and communication systems in terms of digital signal processing as well as hardware architecture provides a realistic possibility for the implementation of integrated radar and communication systems.

The integrated waveform design of Dual-Functional Radar-Communication (DFRC) systems is a primary consideration. At present, integrated waveform design for DFRC systems can be classified into two major categories. The first type of  integrated waveform realizes the information transmission by modulating communication information onto the radar waveform\cite{8917703}. For example, \cite{4268440} designed a linear frequency modulated (LFM) waveform embedded with communication signals, \cite{5545182} developed a theoretical framework for intra-pulse radar-embedded communications and \cite{9449969} proposed an integrated waveform based on frequency-diversity multi-input multi-output (FD-MIMO) by embedding weighted phase-modulated signals into radar waveform and achieved the best tradeoff between location estimation and communication performance.

However, the above integrated waveforms have some disadvantages such as low communication transmission rate and inflexible demodulation methods. To attain high spectral efficiency, the second type of integrated waveform directly uses the communication waveform to complete the detection function. Candidate communication waveforms applicable to detection mainly include spread spectrum signals and orthogonal frequency division multiplexing (OFDM) signals. Among them, spread spectrum signals have good detection ability due to its good correlation properties\cite{507206}. In \cite{7925013}, the integrated waveforms using different spread spectrum sequences were studied and the effects of different kinds of spread spectrum sequences on radar detection performance were compared. \cite{8755489} analyzed the ambiguity function performance of spread spectrum symbol sequences and proposed a symbol sequence optimization method to improve the peak side lobe level. The OFDM waveform inherits the advantages of LFM signal, which has large time-bandwidth product and can improve the distance resolution by joint time-frequency processing. \cite{8059592} analyzed the ambiguity function of the OFDM signals and analysis results indicated that increasing OFDM symbols can resolve the Doppler ambiguity.
\cite{8330462} considered the phase-coded OFDM signals in DFRC system and the proposed methods can achieve higher-resolution range-velocity profiles of the targets
along with the relative high data rate transmissions.
An auto-paired super-resolution range and velocity estimation method in OFDM-DFRC systems was proposed in \cite{9166743}, which improved the resolution of range and velocity estimation without increasing the signal bandwidth and coherent processing interval (CPI).

The combination of OFDM and MIMO technologies can further improve communication and detection performance by effectively utilizing spatial freedom \cite{5422709}, \cite{tse_viswanath_2005}. 
\cite{9562280} designed a hybrid beamforming scheme for wide band MIMO-OFDM DFRC systems, which can achieve the satisfactory radar and communication performance.
\cite{9226446} presented a precoding-based transmitter for MIMO-OFDM DFRC system, where both the communication and radar waveforms occupy the entire available bandwidth at the same time and the BS communicates with multiple downlink users by utilizing the predicted radar interference.
In \cite{9729203}, the radar performance is maximized under constraints of transmit power as well as the communication error rate by selecting the transmit waveforms and the receive filters.

In the aforementioned work, the DFRC systems based on centralized MIMO architecture were mainly considered. Compared with centralized MIMO radar, distributed MIMO systems can further improve the accuracy of target localization \cite{4408448}. \cite{6031934} studied the radar receivers selection in distributed multiple-radar system and proved that same localization accuracy can be achieved by using only a fraction of available radars. In addition, some literatures have investigated the resource allocation in distributed MIMO DFRC systems. \cite{5753953} proposed a power allocation scheme for target localization in distributed multiple-radar systems and showed that significant power savings can be obtained with the proposed strategies. \cite{8835674} studied a narrowband distributed MIMO DFRC system which optimized the power allocation to achieve the desired Cramer-Rao Bound (CRB) of target and communication rate.

In the existing schemes, to the best of our knowledge, the joint optimization of resource allocation and sensors selection in distributed MIMO-OFDM DFRC system was not investigated. Therefore, in this paper, we investigate a distributed dual function MIMO-OFDM system which implements communication and probing in the mode of frequency division. To reduce the feedback overhead and computational complexity, a subset of active radar receivers are selected for processing and feeding back echo signals. Inspired by \cite{9124713}, at the transmitter we design different beamforming vectors on the subcarriers for detection and communication, respectively. Specifically, wide beam design scheme in \cite{4276989} is adopted to guarantee the coverages of detection beams and maximum radio transmission (MRT) is adopted to design beamforming vector for communication. In addition, we derive CRB as detection performance metric instead of transmit beampattern in \cite{9124713}. Compared to \cite{8835674} which only optimized power allocation, the best tradeoff between detection and communication performance is obtained by jointly optimizing power/subcarrier allocation and radar receivers selection. The contributions of this paper are organized as follows.
\begin{itemize}
\item

We propose a frequency division-based communication-sensing fusion scheme in MIMO-OFDM systems. That is, different sets of subcarriers are used for communication and detection respectively. In order to reduce the feedback overhead of echo signals, we restrict the number of active radar receivers. We first determine the beamforming vector of each subcarrier according to the function of the subcarrier, and then analyze the CRB for different detection areas as well as the transmission rate for communication users.

\item

To achieve the best tradeoff between communication rate and detection performance, we formulate a Mixed Integer NonLinear Programming (MINLP) problem which maximizes the total communication rate subject to constraints of detection performance and transmit power. Since the problem is difficult to tackle, we propose an alternative optimization algorithm which solves the original problem by solving two subproblems iteratively. The first subproblem associated with power and subcarriers resource allocation is equivalently converted to be a semidefinite programming (SDP) with difference of convex (DC) approximation. The second subproblem related to radar receivers selection is transformed into a convex quadratic integer problem.

\item
Simulation results show that tradeoff relationship between transmission rate and detection performance with different parameters configuration in distributed MIMO-OFDM DFRC system. The radar performance improvement through radar receivers selection is also demonstrated.  Comparing with existing scheme, our proposed scheme show the advantages in tradeoff performance.
\end{itemize}

The organization of this paper is listed below. Section II
introduces the system model and detection beampattern design, Section III presents the derivation of CRB of detection areas and communication performance. In Section IV, the power/subcarriers allocation and radar receivers selection are jointly optimized to acheive the tradeoff between radar and communication performance. Section V shows the simulation results, and Section VI concludes the paper.

\begin{figure}[h]
\centering
\includegraphics[width=8.5cm,height=7cm] {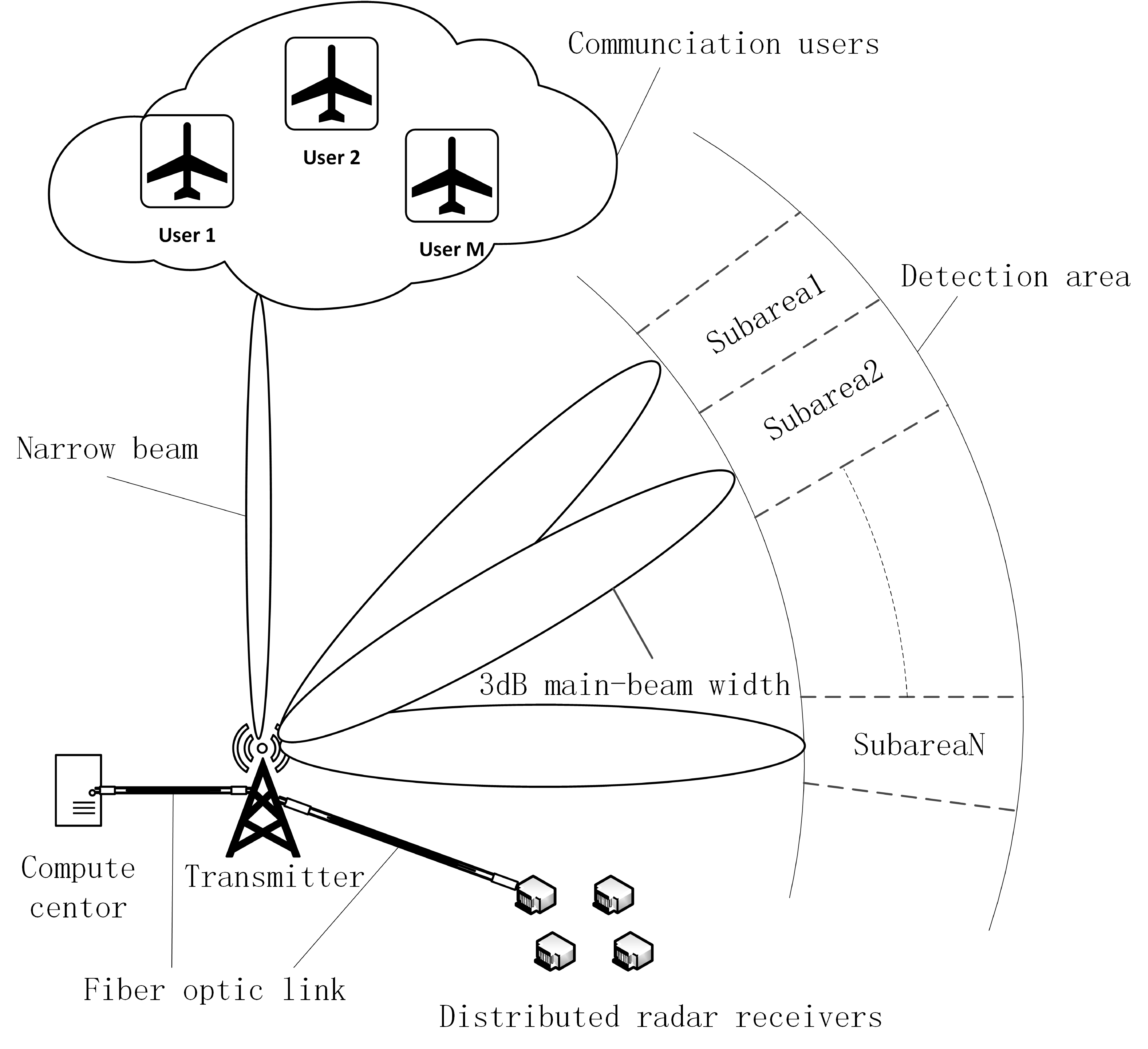}
\caption{MIMO-OFDM based DFRC system model.}
\end{figure}

\section{System Model}
As showed in Fig.1, we consider an integrated MIMO-OFDM system which consists of a transmit BS equipped with $T_x$ antennas, $R_x$ single-antenna radar receivers and a computation center. Suppose that the integrated system serves for $M$ single-antenna communication users and probes an area which is divided into disjoint $N$ parts according to different beam coverages. The transmit BS will simultaneously communicate with communication users and detect targets in specific areas. Echo signals received by the radar receivers are sent back to the computation center for detection and estimation. To reduce the feedback overhead, only the selected radar receivers feed back information to computation center. The set of selected radar receivers is determined by the computation center. In particular, the computing center determines the resource allocation and receiver set based on the channel information between the transmit BS and communication users, as well as the relative position between the detection areas and transmit BS.

In one OFDM symbol duration, each subcarrier may be used to communicate with one of communication users or detect one of beam coverages. Therefore, the baseband signal transmitted on $k$-th subcarrier in the $l$th OFDM symbol is given by
\begin{equation}
	\begin{aligned}
\boldsymbol{s}_{k,l}=\sum_{m=1}^{M}\sqrt{p_k}\sigma_{k,m}^C\boldsymbol{\omega}_{k,m}^Cb_{k,m,l}^C
+\sum_{n=1}^N\sqrt{p_k}\sigma_{k,n}^R\boldsymbol{\Omega}_{k,n}^R \boldsymbol{b}_{k,n,l}^R \label{first}
	\end{aligned}
\end{equation}
where $p_k$ denotes the signal power on the $k$-th subcarrier, $\boldsymbol{\omega}_{k,m}^C \in \boldsymbol{C}^{T_x\times1}$ is the beamforming vector for the $m$-th communication user on the $k$-th subcarrier and ${\boldsymbol{\Omega}_{k,n}^R}\in \boldsymbol{C}^{T_x\times T_x}$ is the beamforming matrix for covering $n$-th detection area on the $k$-th subcarrier.  $\sigma_{k,m}^C $and $\sigma_{k,n}^R$ are subcarrier allocation factors. If $\sigma_{k,m}^C = 1$, the $k$-th subcarrier is allocated to $m$-th communication user; if $\sigma_{k,n}^R = 1$, the $k$-th subcarrier is used to detect the $n$-th subarea. Since one subcarrier could be only assigned to one communication user or one detection subarea, the following condition holds
\begin{equation}
	\begin{aligned}
       \sum_{m=1}^{M}\sigma_{k,m}^C+\sum_{n=1}^N\sigma_{k,n}^R = 1 ,\forall k.
	\end{aligned}
\end{equation}
$b_{k,m,l}^C$ and $\boldsymbol{b}_{k,n,l}^R \in \boldsymbol{C}^{T_x\times1} $ are communication signal with $E(|b_{k,m,l}^C|^2)=1$ and detection signal with $E(\boldsymbol{b}_{k,n,l}^R (\boldsymbol{b}_{k,n,l}^R)^H)=\boldsymbol{I}$ ($\boldsymbol{I}$ is unit matrix), respectively. 
We assume that each OFDM symbol duration is $T_s = T+T_{cp}$ where $T_{cp}$ denotes the length of cyclic prefix (CP), $T=\frac{1}{\Delta f}$ is the OFDM symbol period, and $\Delta f$ is the frequency interval of adjacent sub-channels. Thus, after inverse fast fourier transform (IFFT), adding CP, and impulse sharping, the baseband complex waveform in $l$-th OFDM symbol is given by
\begin{equation}
	\begin{aligned}
\boldsymbol x^{(l)}(t) = \sum_{k=1}^{K}{\boldsymbol s_{k,l}}e^{j2\pi (k-1)\Delta f t}u(t-(l-1)T_s)
	\end{aligned}
\end{equation}
where
\begin{equation}
	u (t) = \left\{
		\begin{aligned}
		& 1,\quad t \in [ -T_{cp}, T ], \\
		& 0,\quad otherwise.
		\end{aligned}	
	\right.
\end{equation}
The bandpass real waveform in $L$ OFDM symbols at carrier frequency $f_c$ is $\boldsymbol x(t) = \text{Re}\left( \sum_{l=1}^{L}\boldsymbol x^{(l)}(t)e^{j2\pi f_c t}\right)$.

In order to maximize receive SNRs at communication terminals, maximum radio transmission (MRT) is adopted to determine $\boldsymbol{\omega}_{k,m}^C$, i.e. $\boldsymbol{\omega}_{k,m}^C = \frac{\boldsymbol{h}_{k,m}}{\Vert\boldsymbol{h}_{k,m}\Vert}$ where $\boldsymbol{h}_{k,m}\in \boldsymbol{C}^{T_x\times 1}$ represents channel vector between BS and the $m$-th communication user on the $k$-th subcarrier. As for detection beams design, to cover the probing area, we generate beampattern with a desired main-beam width on detection subcarriers. The autocorrelated matrix of detection signal on the $k$-th subcarrier for covering $n$-th subarea is $\mathbf{R}_{k,n}=\boldsymbol{\Omega}_{k,n}^R(\boldsymbol{\Omega}_{k,n}^R)^{H}$. The radar beampattern design on the $k$th subcarrier is given by \cite{4276989}
\begin{align}
	&\min_{a,\mathbf{R}_{k,n}}\quad \sum_{q=1}^Q\vert a{P}_n(\theta_q)-\mathbf{a}_k^{H}(\theta_{q})\mathbf{R}_{k,n}\mathbf{a}_k(\theta_q)\vert^{2} \label{M}\\
	 \emph{s.t.}\quad
	& [\mathbf{ R}_{k,n}]_{t,t} = \frac{1}{T_x}, t=1,...,T_x\tag{\ref{M}a}\\
   & \mathbf{ R}_{k,n}\succeq \mathbf{0}, \mathbf{ R}_{k,n}=\mathbf{R}_{k,n}^{H}, a \geq 0\tag{\ref{M}b}
\end{align}
where $a$ is a scaling factor, $\{\theta_q\}_{q=1}^{Q}$ is an equally interval sampling sequence in $[-90^{\circ},90^{\circ}]$,
$Q$ is the number of sampling points, $P_n(\cdot)$ is the desired beampattern for the $n$-th detection area, $\mathbf{a}_k(\theta_q)$ is the steering vector at angle $\theta_q$ on the $k$-th subcarrier and $\mathbf a_k(\theta_{q}) =  \big[ 1,e^{-j\frac{2\pi}{\lambda_k}d \sin \theta_q}, \dots,e^{-j\frac{2\pi}{\lambda_k}\left( T_x - 1\right)d \sin \theta_q} \big] ^{T}$ where $d = \frac{c}{2 f_c}$ is antenna spacing ($c$ denotes the speed of light), $\lambda_k$ is the corresponding wavelength on the $k$-th subcarrier and $\lambda_k = \frac{c}{f_c+(k-1)\Delta f}$. As showed in Fig.1, the detection area is divided into $N$ disjoint subareas. We use ${A}_n$ to denote the angle range in the $n$-th detection subarea, then the $P_n(\cdot)$ in $(\ref{M})$ is designed as
\begin{equation}
	P_n (\theta_{q}) = \left\{
	\begin{aligned}
		& 1,\quad \theta_{q} \in A_n \\
		& 0,\quad otherwise.
	\end{aligned}	
	\right.
\end{equation}
 The problem in $(\ref{M})$ is a semidefinite positive programming (SDP) which can be solved using convex optimization method. We denote the solution of (\ref{M}) as $\mathbf{R}_{k,n}^{*}$. Since $\mathbf{R}_{k,n}^{*}\succeq \mathbf{0}$, we have $\boldsymbol{\Omega}_{k,n}^R$ = $(\mathbf{R}_{k,n}^{*})^{\frac{1}{2}}$.

\section{Analysis of Detection and Communication Performance}
\subsection{Sensing signals design}
To analyze the detection performance in different probing sub-regions, we assume that $N$ targets are located in different sub-regions. We use $\boldsymbol d_{n} = (d_{n}^x,d_{n}^y)$ and $ \boldsymbol{v}_n = (v_{n}^{x},v_{n}^{y})$ to denote the position and the velocity of the target in the $n$-th detection area respectively. In addition, we assume that $\boldsymbol p_0 = (p_0^x,p_0^y)$ and $\boldsymbol p_r = (p_r^x,p_r^y)$ are the positions of the BS and the $r$-th radar receiver respectively. Therefore, we can calculate time delay and Doppler frequency offset from the BS to the $n$-th target then to $r$-th radar receiver as
\begin{equation}
	\begin{aligned}
		\tau_{n,r}={\frac{\Vert\boldsymbol p_0- \boldsymbol d_{n} \Vert+\Vert\boldsymbol d_{n}- \boldsymbol p_r \Vert}{c}}, {\label{CRB1}}
	\end{aligned}
\end{equation}
\vspace{-42pt}
\begin{equation}
	\begin{aligned}
		f_{n,r}={\frac{\boldsymbol{v}_n^T(\boldsymbol{p}_0- \boldsymbol{d}_n)}{\lambda\Vert\boldsymbol{p}_0 - \boldsymbol{d}_n\Vert}}+
              {\frac{\boldsymbol{v}_n^T(\boldsymbol{p}_r- \boldsymbol{d}_n)}{\lambda\Vert\boldsymbol{p}_r - \boldsymbol{d}_n\Vert}} . {\label{CRB2}}
	\end{aligned}
\end{equation}
Then, the equivalent baseband echo signal received by $r$-th radar receiver is
\begin{equation}
	\begin{aligned}
&y_r(t) = \sum_{l=1}^{L}\sum_{n=1}^{N}\sum_{k=1}^{K}s_r c_{n,r}e^{j2\pi f_{n,r}t}\\
&\times  e^{j2\pi (k-1)\Delta f (t-\tau_{n,r}-(l-1)T_s)} u(t-\tau_{n,r}-(l-1)T_s)\\ &\times \boldsymbol{\beta}_{k,n}^H
\sum_{n'=1}^N\sqrt{p_k}\sigma_{k,n'}^R\boldsymbol{\Omega}_{k,n'}^R\boldsymbol{b}_{k,n',l}^R + w(t)
	\end{aligned}
\end{equation}
where $c_{n,r}$ denotes propagation loss from the BS to $n$-th target then to the $r$-th receiver, $w(t)$ is AWGN with zero mean and variance $\sigma_w^{2}$, $\boldsymbol \beta_{k,n}\in \boldsymbol{C}^{T_x\times1}$ is the steering vector on $k$-th subcarrier from the BS to the $n$-th target and $\boldsymbol{\beta}_{k,n} =  \big[ 1,e^{-j\frac{2\pi}{\lambda_k}d \sin \alpha_n}, \dots,e^{-j\frac{2\pi}{\lambda_k}\left( T_x - 1\right)d \sin \alpha_n} \big] ^{T}$, where $\alpha_n = arctan(\frac{d_n^y-p_0^y}{d_n^x-p_0^x}) $ is the angle of departure (AOD) from the BS to the $n$-th target, and binary variable $s_r$ is used to represent the status of the $r$-th radar receiver. Specifically, $s_r=1$ means that the $r$-th radar receiver will feedback echo signals to the computation center, otherwise, it will not be involved in detection. To extract information on the k-th subcarrier, we perform the following transformation on echo signals
\begin{equation}
	\begin{aligned}
		\bar y_{r}(k,l)
		 = {1\over T} \int_{0} ^{T} y_r\left(t +(l-1)T_s\right)e^{-j2\pi (k-1)\Delta ft}dt 	{\label{CRB3}}
	\end{aligned}
\end{equation}
When $\tau_{n,r}\leq T_{cp}$, we can extract information on all subcarriers with (10) to get
\begin{equation}
	\begin{aligned}
		\bar y_{r}(k,l)
		 &= s_r\sum_{n=1}^{N}c_{n,r}e^{j2\pi (l-1)Tf_{n,r}}e^{-j2\pi (k-1)\Delta f\tau_{n,r}}\\
		 &\times
		(\boldsymbol\beta_{k,n}^H\sum_{n'=1}^N\sqrt{p_k}\sigma_{k,n'}^R\boldsymbol{\Omega}_{k,n'}^R\boldsymbol{b}_{k,n',l}^R )+\bar w(k,l),\\
		& \ k=1,\cdots,K {\label{CRB3}}
	\end{aligned}
\end{equation}
where
\begin{equation}
	\begin{aligned}
		\bar w(k,l) = {1\over T} \int_{0} ^{T}w\left(t \right)e^{-j2\pi (k-1)\Delta ft}dt {\label{CRB4}}
	\end{aligned}
\end{equation}
is the AWGN with variance $\sigma_{\bar w}^2 = {\sigma_{w}^2 \over K}$. Based on (11), various estimation algorithms have been proposed \cite{100}, \cite{101}. However, when $\tau_{n,r}> T_{cp}$, the orthogonality between subcarriers is disrupted due to the limitation of $u(t)$ in (3) so that we cannot obtain (11) from (10). Next, we will propose a sensing signal design method so that when $\tau_{n,r}> T_{cp}$, intercarrier-interference (ICI) can be eliminated.

\begin{lemma}
When $\tau_{n,r}> T_{cp}$, we can obtain (11) from (10) if the detection signals satisfy $\boldsymbol b_{k,n,l}^R=e^{j2\pi(k-1)\Delta f T_s}\boldsymbol b_{k,n,l-1}^R$.
\end{lemma}
\noindent\text{Proof}: The proof is given in the Appendix A.

\subsection{CRB analysis}
Based on $(\ref{CRB1})-(\ref{CRB2})$ and $(\ref{CRB3})-(\ref{CRB4})$, we will derive the CRB matrices of position and velocity estimation for all targets. We define unknown parameters $\boldsymbol{\theta} =[(\boldsymbol d_1,\boldsymbol v_1),(\boldsymbol d_2,\boldsymbol v_2),...(\boldsymbol d_N,\boldsymbol v_N)]$ and $\boldsymbol{\varphi}=[(\boldsymbol {\tau}_1,\boldsymbol f_1),(\boldsymbol {\tau}_2,\boldsymbol f_2),...(\boldsymbol {\tau}_N,\boldsymbol f_N)]$, where $ \boldsymbol \tau_{n} = [\tau_{n,1},\tau_{n,2},...,\tau_{n,R_x}], \boldsymbol f_{n} = [f_{n,1},f_{n,2},...,f_{n,R_x}]$. In (\ref{CRB3}), we use $u_r(k,l)$ to denote the mean of $\bar{y}_r(k,l)$  and arrange all $\{\bar{y}_r(k,l)\}$ as a gaussian vector $\boldsymbol{\bar{y}} = [\bar{y}_1(1,1),\bar{y}_1(1,2),$\\$...\bar{y}_{R_x}(K,L)]$ with mean $\boldsymbol{\mu} = [{u}_1(1,1),{u}_1(1,2),...{u}_{R_x}(K,L)]$ and covariance matrix $\boldsymbol E = \sigma_{\bar w}^2\boldsymbol{I}$( $\boldsymbol{I}$ is a $R_x KL \times R_x KL$ unit matrix ). Therefore, we can calculate Fisher Information Matrix (FIM) $\boldsymbol F(\boldsymbol \varphi)$ with respect to $\boldsymbol\varphi$ as
\begin{equation}
	\begin{aligned}
	  \boldsymbol F(\boldsymbol\varphi) = {2\over \sigma_{\bar w}^2}\Re\left(\frac{\partial^{H}\boldsymbol\mu}{\partial\boldsymbol\varphi}\frac{\partial\boldsymbol\mu}{\partial\boldsymbol\varphi}\right) \label{F1}
	\end{aligned}
\end{equation}
where $\Re(x)$ denotes the real part of $x$.
Then according to the chain rule, we can calculate FIM $\boldsymbol F(\boldsymbol \theta)$ with respect to $\boldsymbol\theta$ as
\begin{equation}
	\begin{aligned}
		\boldsymbol F(\boldsymbol\theta) = \boldsymbol{J}\boldsymbol F(\boldsymbol\varphi)\boldsymbol{J}^T \label{F2}
	\end{aligned}
\end{equation}
where $\boldsymbol{J}$ is Jacobian matrix given by
\begin{equation}
	\begin{aligned}
		\boldsymbol{J} = \frac{\partial\boldsymbol{\varphi}}{\partial\boldsymbol{\theta}} =\left[\begin{array}{ccc}
			\boldsymbol{J}_{1} & \dots & \boldsymbol{0}\\
			\vdots & \ddots & \vdots  \\
			\boldsymbol{0} & \dots & \boldsymbol{J}_{N}
		\end{array}\right] \label{F3}
	\end{aligned}
\end{equation}
where
\begin{equation}
	\begin{aligned}
	\boldsymbol{J}_{n}=\left[\begin{array}{cccccc}
		\frac{\partial{\tau}_{n,1}}{\partial{d}_{n}^{x}} & \dots&\frac{\partial{\tau}_{n,R_x}}{\partial{d}_{n}^{x}} &\frac{\partial{f}_{n,1}}{\partial{d}_{n}^{x}}& \dots&\frac{\partial{f}_{n,R_x}}{\partial{d}_{n}^{x}} \\
		\frac{\partial{\tau}_{n,1}}{\partial{d}_{n}^{y}} & \dots&\frac{\partial{\tau}_{n,R_x}}{\partial{d}_{n}^{y}} &\frac{\partial{f}_{n,1}}{\partial{d}_{n}^{y}}& \dots&\frac{\partial{f}_{n,R_x}}{\partial{d}_{n}^{y}} \\
		0 & \dots&0 &\frac{\partial{f}_{n,1}}{\partial{v}_{n}^{x}}& \dots&\frac{\partial{f}_{n,R_x}}{\partial{v}_{n}^{x}} \\
		0 & \dots&0 &\frac{\partial{f}_{n,1}}{\partial{v}_{n}^{y}}& \dots&\frac{\partial{f}_{n,R_x}}{\partial{v}_{n}^{y}}\end{array}\right]  \label{F4}.
	\end{aligned}
\end{equation}
Based on $(\ref{F1})$-$(\ref{F2})$, $\boldsymbol{F}(\boldsymbol{\theta})$ can be represented in the following form
\begin{equation}
	\begin{aligned}
		\boldsymbol{F}(\boldsymbol{\theta})= \left[\begin{array}{cccc}
			\boldsymbol{F}_{1}^{(1)} & \boldsymbol{F}_{2}^{(1)} & \dots & \boldsymbol{F}_{N}^{(1)} \\
			\boldsymbol{F}_{1}^{(2)} & \boldsymbol{F}_{2}^{(2)} & \dots & \boldsymbol{F}_{N}^{(2)}  \\
			\vdots & \vdots &  \ddots & \vdots \\
			\boldsymbol{F}_{1}^{(N)} & \boldsymbol{F}_{2}^{(N)} & \dots & \boldsymbol{F}_{N}^{(N)}
		\end{array}\right]
	\end{aligned}
\end{equation}
where $\boldsymbol F_i^{(j)}$s are $4\times 4$ matrices.
Diagonal elements of $\boldsymbol{F}^{-1}(\boldsymbol{\theta})$ are CRB of estimating $\boldsymbol{\theta}$. However, accurate expressions of CRB are too complicated to be applied to the optimization problem in Sec. IV. It is observed that non-diagonal blocks of  $\boldsymbol{F}^{-1}(\boldsymbol{\theta})$ denote coupling between targets. When targets are far apart, elements in non-diagonal block will be very small. On the other hand, even when targets are close, we can get the lower bound of CRB by ignoring the non-diagonal blocks \cite{}. Therefore, in the following analysis, we only consider the following block diagonal matrix
\begin{equation}
	\begin{aligned}
		\boldsymbol{\bar{F}}(\boldsymbol{\theta}) =\left[\begin{array}{ccc}\label{xx}
			\boldsymbol{F}_{1}^{(1)} & \dots & \boldsymbol{0}  \\
			\vdots & \ddots & \vdots  \\
			\boldsymbol{0} & \dots & \boldsymbol{F}_{N}^{(N)}
		\end{array}\right].
	\end{aligned}
\end{equation}
The difference between accurate CRB and lower bound of CRB is exhibited in simulation results. By ignoring the coupling between location and velocity estimation, $\boldsymbol{F}_{n}^{(n)}$ has the form of
\begin{equation}
	\begin{aligned}
	  \boldsymbol{F}_{n}^{(n)}=\sum_{r=1}^{R_x}\sum_{k=1}^{K}\sum_{n'=1}^{N}s_r p_k\sigma_{k,n'}^R  \boldsymbol{G}_{n}(r,k,n') \label{F}
	\end{aligned}
\end{equation}
where $\boldsymbol{G}_{n}(r,k,n')$ is a $4 \times 4$ block diagonal matrix
\begin{equation}
	\begin{aligned}
		\boldsymbol{G}_{n}(r,k,n') =\left[\begin{array}{cc}
			\boldsymbol{D}_{n}(r,k,n') & \boldsymbol{0}  \\
			\boldsymbol{0}  & \boldsymbol{V}_{n}(r,k,n')
		\end{array}\right]
	\end{aligned}
\end{equation}
where both $\boldsymbol{D}_{n}(r,k,n')$ and $\boldsymbol{V}_{n}(r,k,n')$ are $2\times 2$  matrices, and their expressions are given in the Appendix A. Based on [17] we can obtain the CRB matrix of the $n$-th target $\boldsymbol{C}_{n}= ({\boldsymbol{F}_{n}^{(n)}})^{-1}$. To facilitate later analysis, we use $\boldsymbol{C}_{n}^{d}$ and $\boldsymbol{C}_{n}^{v}$ to denote CRB matrices of location and velocity respectively, which are given by
\begin{equation}
	\begin{aligned}
	  \boldsymbol{C}_{n}^{d} = \left(\sum_{r=1}^{R_x}\sum_{k=1}^{K}\sum_{n'=1}^{N}s_r p_k\sigma_{k,n'}^R  \boldsymbol{D}_{n}(r,k,n')\right)^{-1}\label{Cd}
	\end{aligned}
\end{equation}
\begin{equation}
	\begin{aligned}
	  \boldsymbol{C}_{n}^{v} = \left(\sum_{r=1}^{R_x}\sum_{k=1}^{K}\sum_{n'=1}^{N}s_r p_k\sigma_{k,n'}^R  \boldsymbol{V}_{n}(r,k,n')\right)^{-1}\label{Cv}.
	\end{aligned}
\end{equation}

\subsection{Communication rate}
In this subsection, we analyze the communication performance of the integrated system. If the $k$-th subcarrier is allocated to the $m$-th communication user, then the received signal of the $m$-th communication user on the $k$-th subcarrier is
\begin{equation}
	\begin{aligned}
	  r_{k,m} = \boldsymbol h_{k,m}^{H}\sqrt{p_k}\boldsymbol{\omega}_{k,m}^C b_m^C+z {\label{noise_c}}
	\end{aligned}
\end{equation}
where $\boldsymbol{h}_{k,m} = a_m\mathbf a_k(\gamma_{m})$ is the channel vector between the BS and the $m$-th communication user on the $k$-th subcarrier, $a_m$ represents the propagation loss from the BS to the $m$-th communication user, $\mathbf a_k(\gamma_{m}) =  \big[ 1,e^{-j\frac{2\pi}{\lambda_k}d \sin \gamma_m}, \dots,e^{-j\frac{2\pi}{\lambda_k}\left( T_x - 1\right)d \sin \gamma_m} \big] ^{T}$ and $\gamma_m$ denotes the angle from the BS to $m$-th communication user, $z$ is AWGN with zero mean and variance  $\sigma_z^2$.
When $ \boldsymbol\omega_{k,m}^C = \frac{\boldsymbol{h}_{k,m}}{\Vert\boldsymbol{h}_{k,m}\Vert} $, the transmission rate of the $m$-th communication user on the $k$-th subcarrier is
\begin{equation}
	\begin{aligned}
		R_{k,m} = \log\Big(1+\frac{a_m^2\Vert\boldsymbol{h}_{k,m}\Vert^2 p_k}{\sigma_{z}^2}\Big). \label{calR}
	\end{aligned}
\end{equation}

\section{Resource Allocation and Receivers Selection}
In this section, we investigate how to maximize the transmission rate under the constraints of CRB and total transmitted power by optimizing the $\{\sigma_{k,n}^{R}\},\{\sigma_{k,m}^{C}\},\{p_k\}$ and $\{s_r\}$. The optimization problem is formulated as
\begin{align}
	 \max&_{\{\sigma_{k,n}^{R}\},\{\sigma_{k,m}^{C}\},\{p_k\} ,\{s_r\}} \sum_{k=1}^{K} \sum_{m=1}^{M} \sigma_{k,m}^{C}R_{k,m}\label{P} \\
	\emph{s.t.} \quad
 	& [\boldsymbol{C}_{n}^d]_{1,1}\leq\eta_d ,[\boldsymbol{C}_{n}^d]_{2,2}\leq\eta_d, \forall {n}\tag{\ref{P}a}\\
	& [\boldsymbol{C}_{n}^v]_{1,1}\leq\eta_v ,[\boldsymbol{C}_{n}^v]_{2,2}\leq\eta_v, \forall {n} \tag{\ref{P}b}\\
   & \sum_{m=1}^{M}\sigma_{k,m}^{C}+\sum_{n=1}^{N}\sigma_{k,n}^{R}=1,\forall k\tag{\ref{P}c}\\
   & \sum_{k=1}^{K}p_k\leq P_{max}\tag{\ref{P}d}\\
	& \sum_{r=1}^{R_x}s_r = N_r \tag{\ref{P}e}.
\end{align}
$(\ref{P}a)$ and $(\ref{P}b)$ are the estimation performance constraints, where $\eta_d$ and $\eta_v$ denote the CRB bounds of distance and velocity respectively. $(\ref{P}c)$ means that each subcarrier can only be allocated to one communication user or one detection subarea. $(\ref{P}d)$ is the total transmit power constraint. $(\ref{P}e)$ is the constraint of the number of available radar receivers, i.e. $N_r\leq R_x$. Because $(\ref{P}a)$  and $(\ref{P}b)$ are nonlinear constraints which contains both continuous and discrete variables, the problem $(\ref{P})$ is a Mixed Integer NonLinear Programming (MINLP) problem.

We propose an alternative optimization method to solve $(\ref{P})$ which divides $(\ref{P})$ into two subproblems, denoted by $(\ref{A})$ and $(\ref{B})$, respectively. For given radar receiver set, we solve $(\ref{A})$ to obtain the optimal subcarrier and power allocation,
\begin{align}
	 \max&_{\{\sigma_{k,m}^{C}\},\{\sigma_{k,n}^{R}\},\{p_k\}}\sum_{k=1}^{K} \sum_{m=1}^{M} \sigma_{k,m}^C R_{k,m}\label{A} \\
	\emph{s.t.}: \quad
	&[\boldsymbol{C}_{n}^d]_{1,1}\leq\eta_d ,[\boldsymbol{C}_{n}^d]_{2,2}\leq\eta_d, \forall {n}\tag{\ref{A}a}\\
	& [\boldsymbol{C}_{n}^v]_{1,1}\leq\eta_v ,[\boldsymbol{C}_{n}^v]_{2,2}\leq\eta_v, \forall {n} \tag{\ref{A}b}\\
   & \sum_{m=1}^{M}\sigma_{k,m}^{C}+\sum_{n=1}^{N}\sigma_{k,n}^{R}=1,\forall k\tag{\ref{A}c}\\
   &\sum_{k=1}^{K}p_k\leq P_{max}\tag{\ref{A}d}
\end{align}
Then, for given subcarrier/power allocation, we solve $(\ref{B})$ to get the optimal radar receiver set,
\begin{align}
	 \min&_{\{s_r, \eta_d\}}\quad \eta_d   \label{B} \\
	 \emph{s.t.}:\quad
	&  [\boldsymbol{C}_{n}^d]_{1,1}\leq\eta_d, [\boldsymbol{C}_{n}^d]_{2,2}\leq\eta_d, \forall {n}\tag{\ref{B}a}\\
	&  [\boldsymbol{C}_{n}^v]_{1,1}\leq\eta_v , [\boldsymbol{C}_{n}^v]_{2,2}\leq\eta_v, \forall {n} \tag{\ref{B}b}\\
	& \sum_{r=1}^{R_x}s_r = N_r \tag{\ref{B}c}&
\end{align}
or
\begin{align}
	 \min&_{\{s_r, \eta_v\}}\quad  \eta_v \label{B1} \\
	 \emph{s.t.}:\quad
	&  [\boldsymbol{C}_{n}^d]_{1,1}\leq\eta_d, [\boldsymbol{C}_{n}^d]_{2,2}\leq\eta_d, \forall {n}\tag{\ref{B1}a}\\
	&  [\boldsymbol{C}_{n}^v]_{1,1}\leq\eta_v , [\boldsymbol{C}_{n}^v]_{2,2}\leq\eta_v, \forall {n} \tag{\ref{B1}b}\\
	& \sum_{r=1}^{R_x}s_r = N_r. \tag{\ref{B1}c}&
\end{align}
In $(\ref{P})$, when subcarrier/power allocation is given, the objective function is independent of $\{s_r\}$. However, to improve the estimation performance, in $(\ref{B})$ and $(\ref{B1})$, we set $\eta_d$ and $\eta_v$ as the objective function, and optimize $\{s_r\}$ to minimize $\eta_d$ or $\eta_v$. When we intend to minimize the CRB of target position, $\eta_v$ is set to a constant, and ({\ref{P}}) is solved by solving ({\ref{A}}) and ({\ref{B}}) iteratively. Likewise, if we intend to minimize the CRB of target velocity, $\eta_d$ is set to a constant, and then ({\ref{P}}) is solved by solving ({\ref{A}}) and ({\ref{B1}}). Due to the similarity between the problem (\ref{B}) and the problem (\ref{B1}), in the following we only consider the solution to subproblem (\ref{B}) below.

\subsection{Subcarriers and Power Allocation}

In this subsection, we mainly solve the subproblem in ($\ref{A}$). Firstly, we relax the binary variables $\sigma_{k,m}^{C}$ and $\sigma_{k,n}^{R}$ to real number in $(0,1]$ and make variables transformation: $\bar{p}_{k,m}^{C} = p_k\sigma_{k,m}^{C}$, $\bar{p}_{k,n}^{R} = p_k\sigma_{k,n}^{R}$. The objective function in ($\ref{A}$) is transformed to
\begin{equation}
	\begin{aligned}
		\sum_{k=1}^K\sum_{m=1}^M{\bar R}_{k,m} = \sum_{k=1}^K\sum_{m=1}^M\sigma_{k,m}^{C}\log\Big(1+\frac{a_m^{2}\Vert\boldsymbol{h}_{k,m}\Vert^2 \bar{p}_{k,m}^{C}}{\sigma_{\bar z}^2\sigma_{k,m}^{C}}\Big) \label{R}
	\end{aligned}
\end{equation}
which is a concave function of $(\sigma_{k,m}^{C},\bar{p}_{k,m}^{C})$. Although $\sigma_{k,m}^{C}$ in (\ref{R}) is not allowed to be 0, when $\sigma_{k,m}^C$ is arbitraily close to 0,  ${\bar R}_{k,m} $ also approaches 0.
Next, we prove that $({\ref{A}a})$ and $({\ref{A}b})$ can be equivalently converted to convex constraints.
According to expressions in (20), $[\boldsymbol{C}_{n}^d]_{1,1}\leq\eta_d$ can be written in the following form
\begin{equation}
	\begin{aligned}
	\bigg[\left(\begin{array}{cc} \label{trans}
			\boldsymbol{a}^{T} \boldsymbol{x} & \boldsymbol{b}^{T} \boldsymbol{x} \\
			\boldsymbol{b}^{T} \boldsymbol{x} & \boldsymbol{d}^{T} \boldsymbol{x}
		\end{array}\right)^{-1}\bigg]_{1,1}\leq \eta
	\end{aligned}
\end{equation}
where $\boldsymbol{a},\boldsymbol{b},\boldsymbol{d}$ are constant value vectors and $\boldsymbol{x}$ is a variable. Then, we have
\begin{equation}
\begin{aligned}
		\frac{\boldsymbol{d}^T\boldsymbol{x}}{\boldsymbol{x}^{T}(\boldsymbol{a}\boldsymbol{d}^{T} - \boldsymbol{b}\boldsymbol{b}^T)\boldsymbol{x}} \leq \eta
	 \Longleftrightarrow &(\eta\boldsymbol{a}^{T}\boldsymbol{x}-1)\boldsymbol{d}^{T}\boldsymbol{x} \\&- (\sqrt{\eta}\boldsymbol{b}^{T}\boldsymbol{x})^2 \geq 0 \label{conexp2}
\end{aligned} 
\end{equation}
Construct matrix
$\boldsymbol A=\left[{\begin{array}{cc}\boldsymbol{d}^{T}\boldsymbol{x} &  \sqrt{\eta}\boldsymbol{b}^{T}\boldsymbol{x} \\
			\sqrt{\eta}\boldsymbol{b}^{T}\boldsymbol{x} & \eta\boldsymbol{a}^{T}\boldsymbol{x}-1 \end{array}}
			\right] $. Because $\boldsymbol{d}^{T}\boldsymbol{x} >0$ and (29), we have $\boldsymbol A \succeq  \boldsymbol{0}$, which is a convex constraint of $\boldsymbol{x}$. Similarly, all constraints in (25a) and (25b) can be transformed to the convex constraints in (30)-(33). For notational simplicity,  we use $\boldsymbol{D}_{n}$ and $\boldsymbol{V}_{n}$ to represent $\boldsymbol{D}_{n}(r,k,n')$ and $\boldsymbol{V}_{n}(r,k,n')$ respectively, and $\sum_{r=1}^{Rx}\sum_{k=1}^{K}\sum_{n'=1}^{N}(\cdot)$ is abbreviated as $\sum_{r,k,n'}(\cdot)$.

\begin{equation}
	\begin{aligned}
	&\boldsymbol{\bar{C}}_{n}^{(d,1)} = \label{CE1}\\ &
	\left[\begin{array}{cc}
			\sum\limits_{r,k,n'} s_r\bar{p}_{k,n'}^R [\boldsymbol{D}_{n}]_{2,2} & \sqrt{\eta_d}\sum\limits_{r,k,n'} s_r\bar{p}_{k,n'}^R   [\boldsymbol{D}_{n}]_{1,2}  \\
			\sqrt{\eta_d}\sum\limits_{r,k,n'} s_r\bar{p}_{k,n'}^R  [\boldsymbol{D}_{n}]_{2,1} &\eta_d\sum\limits_{r,k,n'} s_r\bar{p}_{k,n'}^R [\boldsymbol{D}_{n}]_{1,1}-1
		\end{array}\right]
	\end{aligned}
\end{equation}
\begin{equation}
	\begin{aligned}
	&\boldsymbol{\bar{C}}_{n}^{(d,2)} =\\
	&\left[\begin{array}{cc}
			\sum\limits_{r,k,n'} s_r\bar{p}_{k,n'}^R  [\boldsymbol{D}_{n}]_{1,1} & \sqrt{\eta_d}\sum\limits_{r,k,n'} s_r\bar{p}_{k,n'}^R   [\boldsymbol{D}_{n}]_{1,2}  \\
			\sqrt{\eta_d}\sum\limits_{r,k,n'} s_r\bar{p}_{k,n'}^R  [\boldsymbol{D}_{n}]_{2,1}  & \eta_d\sum\limits_{r,k,n'} s_r\bar{p}_{k,n'}^R [\boldsymbol{D}_{n}]_{2,2}-1
		\end{array}\right]
	\end{aligned}
\end{equation}
\begin{equation}
	\begin{aligned}
	&\boldsymbol{\bar{C}}_{n}^{(v,1)} =\\
	&\left[\begin{array}{cc}
			\sum\limits_{r,k,n'} s_r\bar{p}_{k,n'}^R  [\boldsymbol{V}_{n}]_{2,2} & \sqrt{\eta_v}\sum\limits_{r,k,n'} s_r\bar{p}_{k,n'}^R   [\boldsymbol{V}_{n}]_{1,2}  \\
			\sqrt{\eta_v}\sum\limits_{r,k,n'} s_r\bar{p}_{k,n'}^R  [\boldsymbol{V}_{n}]_{2,1}  & \eta_v\sum\limits_{r,k,n'} s_r\bar{p}_{k,n'}^R [\boldsymbol{V}_{n}]_{1,1}-1
		\end{array}\right]
	\end{aligned}
\end{equation}
\begin{equation}
	\begin{aligned}
	&	\boldsymbol{\bar{C}}_{n}^{(v,2)} = \label{CE4}
	\\&\left[\begin{array}{cc}
			\sum\limits_{r,k,n'} s_r\bar{p}_{k,n'}^R  [\boldsymbol{V}_{n}]_{1,1} & \sqrt{\eta_v}\sum\limits_{r,k,n'} s_r\bar{p}_{k,n'}^R   [\boldsymbol{V}_{n}]_{1,2}  \\
			\sqrt{\eta_v}\sum\limits_{r,k,n'} s_r\bar{p}_{k,n'}^R  [\boldsymbol{V}_{n}]_{2,1}  & \eta_v\sum\limits_{r,k,n'} s_r\bar{p}_{k,n'}^R [\boldsymbol{V}_{n}]_{2,2}-1
		\end{array}\right]
	\end{aligned}
\end{equation}
 Then, we obtain the following convex problem
\begin{align}
	 \max&_{\{\sigma_{k,m}^{C}\},\{\sigma_{k,n}^{R}\},\{\bar p_{k,m}^C\},\{\bar p_{k,n}^R\}} \sum_{k=1}^{K} \sum_{m=1}^{M} \bar {R}_{k,m}\label{A'} \\
	\emph{s.t.}:
	& \quad\boldsymbol{\bar{C}}_{n}^{(d,1)}\succeq \boldsymbol{0}, \boldsymbol{\bar{C}}_{n}^{(d,2)}\succeq \boldsymbol{0}, \forall {n}\tag{\ref{A'}a}\\
	& \quad \boldsymbol{\bar{C}}_{n}^{(v,1)}\succeq \boldsymbol{0}, \boldsymbol{\bar{C}}_{n}^{(v,2)}\succeq \boldsymbol{0}, \forall {n} \tag{\ref{A'}b}\\
   & \sum_{m=1}^{M}\sigma_{k,m}^{C}+\sum_{n=1}^{N}\sigma_{k,n}^{R}=1,\forall k\tag{\ref{A'}c}\\
   & \sum_{k=1}^{K}\Big(\sum_{m=1}^{M}\bar{p}_{k,m}^C+\sum_{n=1}^{N}\bar{p}_{k,n}^R\Big)\leq P_{max}\tag{\ref{A'}d}.&
\end{align}
However, the solution to (\ref{A'}) can not guarantee that the relaxed variables $\{\sigma_{k,n}^R\}$ and $\{\sigma_{k,m}^C\}$ will value at the endpoints of interval $(0,1]$. To address this issue, we introduce the constraints (\ref{C1}) and (\ref{C2}) to restrict the value of $\{\sigma_{k,n}^R\}$ and $\{\sigma_{k,m}^C\}$ near the endpoints of interval $(0,1]$,
\begin{equation}
	\begin{aligned}
		\sigma_{k,n}^R(1-\sigma_{k,n}^R)\leq \alpha_{k,n}\quad\forall k,n \label{C1}
	\end{aligned}
\end{equation}
\begin{equation}
	\begin{aligned}
		\sigma_{k,m}^C(1-\sigma_{k,m}^C)\leq \delta_{k,m}\quad\forall k,m \label{C2}
	\end{aligned}
\end{equation}
where $\{\alpha_{k,n}\}$ and $\{\delta_{k,m}\}$ are the slack variables. Since (\ref{C1}) and (\ref{C2}) have the forms of Difference of Convex (DC) programming, we adopt the following first-order Taylor series at given points $\{\sigma_{k,n}^{R,(j)}\}$ and  $\{\sigma_{k,m}^{C,(j)}\}$ to approximate them,
\begin{equation}
	\begin{aligned}
		(\sigma_{k,n}^{R,(0)})^2 + \sigma_{k,n}^R(1-2\sigma_{k,n}^{R,(0)})\leq \alpha_{k,n} ,\quad\forall k,n \label{C3}
	\end{aligned}
\end{equation}
\begin{equation}
	\begin{aligned}
		(\sigma_{k,m}^{C,(0)})^2 + \sigma_{k,m}^C(1-2\sigma_{k,m}^{C,(0)})\leq \delta_{k,m} .\quad\forall k,m\label{C4}
	\end{aligned}
\end{equation}
By adding $(\ref{C3})$ and $(\ref{C4})$, the problem $(\ref{A'})$ is further converted to
\begin{align}
\max&_{\{\sigma_{k,m}^{C}\},\{\sigma_{k,n}^{R}\},\{\bar p_{k,m}^C\},\{\bar p_{k,n}^R\},\{\alpha_{k,n}\},\{\delta_{k,m}\}}\bar{R}-\beta^{(j)} P \label{A''}\\
	\emph{s.t.}:
	 &\quad\boldsymbol{\bar{C}}_{n}^{(d,1)}\succeq \boldsymbol{0}, \boldsymbol{\bar{C}}_{n}^{(d,2)}\succeq \boldsymbol{0}, \forall {n}  \tag{\ref{A''}a}\\
	& \quad \boldsymbol{\bar{C}}_{n}^{(v,1)}\succeq \boldsymbol{0}, \boldsymbol{\bar{C}}_{n}^{(v,2)}\succeq \boldsymbol{0}, \forall {n} \tag{\ref{A''}b} \\
   & \sum_{m=1}^{M}\sigma_{k,m}^{C}+\sum_{n=1}^{N}\sigma_{k,n}^{R}=1,\forall k \tag{\ref{A''}c}\\
   &\sum_{k=1}^{K}\Big(\sum_{m=1}^{M}\bar{p}_{k,m}^C+\sum_{n=1}^{N}\bar{p}_{k,n}^R\Big)\leq P_{max} \tag{\ref{A''}d}\\
	&(\sigma_{k,n}^{R,(j)})^2 + \sigma_{k,n}^R(1-2\sigma_{k,n}^{R,(j)})\leq \alpha_{k,n} \quad\forall k,n \tag{\ref{A''}e}\\
	&(\sigma_{k,m}^{C,(j)})^2 + \sigma_{k,m}^C(1-2\sigma_{k,m}^{C,(j)})\leq \delta_{k,m} \quad\forall k,m \tag{\ref{A''}f} &
\end{align}
where $\bar{R}=\sum_{k=1}^{K} \sum_{m=1}^{M} \bar {R}_{k,m}$, $\beta^{(j)}>0$ denotes the penalty coefficient at the $j$-th iteration, $ P = \sum_{k=1}^{K}\sum_{n=1}^{N}\alpha_{k,n}+ \sum_{k=1}^{K}\sum_{m=1}^{M}\delta_{k,m}$ denotes the total violation, $\sigma_{k,n}^{R,(j)}$ and $\sigma_{k,m}^{C,(j)}$ are optimal solution at the $j$-th iteration. We state the algorithm to solve problem (\ref{A'}) in Algorithm 1.
\begin{algorithm}
    \caption{ Iterative Algorithm for Solving Problem (\ref{A'}) }
    \label{alg:2}
    \begin{algorithmic}
         \REQUIRE initial points $\{\sigma_{k,n}^{R,(0)}\}$, $\{\sigma_{k,m}^{C,(0)}\}$,  maximum penalty $\beta_{max}$, intital penalty $\beta^{(0)}$, a constant $\gamma>1$, the iteration number $j = 0$, the tolerance $\epsilon$.
        \REPEAT
        \STATE obtain the solution $\{\sigma_{k,n}^{R,(*)}\}$,$\{\sigma_{k,m}^{C,(*)}\}$,$\{\bar{p}_{k,n}^{R,(*)}\}$,$\{\bar{p}_{k,m}^{C,(*)}\}$ by solving problem  $(\ref{A''})$;
        \STATE set $j = j+1$;
 		  \STATE update $\sigma_{k,n}^{R,(j)}=\sigma_{k,n}^{R,(*)}$ and $\sigma_{k,m}^{C,(j)}=\sigma_{k,m}^{C,(*)}$;
			\STATE $\beta^{(j)} = \min\{ \gamma\beta^{(j-1)}, \beta_{max} \}$;
        \UNTIL $|R^{(j)}-R^{(j-1)}|\leq \epsilon $ .
	    \ENSURE $\{\sigma_{k,n}^{R,(*)}\}$,$\{\sigma_{k,m}^{C,(*)}\}$,$\{\bar{p}_{k,n}^{R,(*)}\}$,$\{\bar{p}_{k,m}^{C,(*)}\}$.
    \end{algorithmic}
\end{algorithm}

\subsection{Radar Receivers Selection}
In this subsection, we mainly tackle the subproblem in $(\ref{B})$. We first introduce a Lemma for later use.
\begin{lemma}
  Suppose that $a_i\in \boldsymbol R$ and $\boldsymbol B_i$ is $2\times2$ matrix, $i = 1,2,...,I$. We denote $\boldsymbol{a}=(a_1,a_2,...a_I)^{T}$. Then the following equations hold
\begin{equation}
	\begin{aligned}
	\left[(\sum_{i=1}^{I}a_i\boldsymbol B_i)^{-1}\right]_{1,1}=\frac{\boldsymbol{a}^{T}\boldsymbol{p}_1}{\boldsymbol{a}^{T}\boldsymbol{Q}\boldsymbol{a}}
	\end{aligned}
\end{equation}

\begin{equation}
	\begin{aligned}
	\left[(\sum_{i=1}^{I}a_i\boldsymbol B_i)^{-1}\right]_{2,2}=\frac{\boldsymbol{a}^{T}\boldsymbol{p}_2}{\boldsymbol{a}^{T}\boldsymbol{Q}\boldsymbol{a}}
	\end{aligned}
\end{equation}
\end{lemma}
\noindent where $\boldsymbol Q$ is a $I \times I$ matrix with $[\boldsymbol Q]_{i,i'}=\left[\boldsymbol B_i\right]_{1,1}\left[\boldsymbol B_{i'}\right]_{2,2}-\left[\boldsymbol B_{i}\right]_{1,2}\left[\boldsymbol B_{i'}\right]_{2,1}$, $\boldsymbol{p}_1 = ([\boldsymbol B_1]_{2,2},[\boldsymbol B_2]_{2,2},...,[\boldsymbol B_I]_{2,2})^{T}$ and $\boldsymbol{p}_2 = ([\boldsymbol B_1]_{1,1},[\boldsymbol B_2]_{1,1},...,[\boldsymbol B_I]_{1,1})^{T}$.

\text{Proof}: The proof is given in the Appendix D.

To tackle $(\ref{B}a)$ and $(\ref{B}b)$ with Lemma 2, we first define
\begin{equation}
	\begin{aligned}
	\boldsymbol B_{n,r}^{d}=\sum_{k=1}^{K}\sum_{n'=1}^{N}p_k^{R}\sigma_{k,n'}^{R}\boldsymbol D_{n}(r,k,n')\\
	\end{aligned}
\end{equation}
\begin{equation}
	\begin{aligned}
	\boldsymbol B_{n,r}^{v}=\sum_{k=1}^{K}\sum_{n'=1}^{N}p_k^{R}\sigma_{k,n'}^{R}\boldsymbol V_{n}(r,k,n').
	\end{aligned}
\end{equation}
Then, according to Lemma 2, $(\ref{B}a)$ and $(\ref{B}b)$ can be expressed as
\begin{equation}
	\begin{aligned}
	{[}\boldsymbol{C}_{n}^{d}{]}_{1,1}=\big[(\sum_{r=1}^{R_x}s_r\boldsymbol{B}_{n,r}^{d})^{-1}\big]_{1,1}=\frac{\boldsymbol{s}^{T}\boldsymbol{p}_{n}^{d,1}}
	{\boldsymbol{s}^{T}\boldsymbol{Q}_{n}^{d}\boldsymbol{s}}\label{deno1}
	\end{aligned}
\end{equation}

\begin{equation}
	\begin{aligned}
		{[}\boldsymbol C_{n}^d{]}_{2,2} = \Big[(\sum_{r=1}^{R_x}s_r\boldsymbol B_{n,r}^{d})^{-1}\Big]_{2,2}=\frac{\boldsymbol{s}^{T}\boldsymbol {p}_{n}^{d,2}}{\boldsymbol{s}^{T}\boldsymbol {Q}_{n}^{d}\boldsymbol{s}}\label{deno2}
	\end{aligned}
\end{equation}

\begin{equation}
	\begin{aligned}
	{[}\boldsymbol C_{n}^v{]}_{1,1} = \Big[(\sum_{r=1}^{R_x}s_r\boldsymbol B_{n,r}^{v})^{-1}\Big]_{1,1}=\frac{\boldsymbol{s}^{T}\boldsymbol {p}_{n}^{v,1}}{\boldsymbol{s}^{T}\boldsymbol {Q}_{n}^{v}\boldsymbol{s}}\label{deno3}
	\end{aligned}
\end{equation}
\begin{equation}
	\begin{aligned}
	{[}\boldsymbol C_{n}^v{]}_{2,2} = \Big[(\sum_{r=1}^{R_x}s_r\boldsymbol B_{n,r}^{v})^{-1}\Big]_{2,2}=\frac{\boldsymbol{s}^{T}\boldsymbol {p}_{n}^{v,2}}{\boldsymbol{s}^{T}\boldsymbol {Q}_{n}^{v}\boldsymbol{s}}\label{deno4}
	\end{aligned}
\end{equation}
where $\boldsymbol {Q}_n^{d}$ and $\boldsymbol {Q}_n^{v}$ are $R_x\times R_x$ matrices with
$[\boldsymbol {Q}_n^{d}]_{r,r'}=[\boldsymbol B_{n,r}^{d}]_{1,1}[\boldsymbol B_{n,r'}^{d}]_{2,2}-[\boldsymbol B_{n,r}^{d}]_{1,2}[\boldsymbol B_{n,r'}^{d}]_{2,1}$,
$[\boldsymbol {Q}_n^{v}]_{r,r'}=\left[\boldsymbol B_{n,r}^{v}]_{1,1}[\boldsymbol B_{n,r'}^{v}\right]_{2,2}-[\boldsymbol B_{n,r}^{v}]_{1,2}[\boldsymbol B_{n,r'}^{v}]_{2,1}$, 
$\boldsymbol p_{n}^{d,1}= ([\boldsymbol B_{n,1}^{d}]_{2,2},[\boldsymbol B_{n,2}^{d}]_{2,2},...,[\boldsymbol B_{n,R_x}^{d}]_{2,2})^{T}$,

\noindent $\boldsymbol p_{n}^{d,2} = ([\boldsymbol B_{n,1}^{d}]_{1,1},[\boldsymbol B_{n,2}^{d}]_{1,1},...,[\boldsymbol B_{n,R_x}^{d}]_{1,1})^{T}$, 
$\boldsymbol p_{n}^{v,1} = ([\boldsymbol  B_{n,1}^{v}]_{2,2},[\boldsymbol  B_{n,2}^{v}]_{2,2},...,[\boldsymbol  B_{n,R_x}^{v}]_{2,2})^{T}$, $\boldsymbol p_{n}^{v,2} = ([\boldsymbol  B_{n,1}^{v}]_{1,1},$\\
 $[\boldsymbol  B_{n,2}^{v}]_{1,1},...,[\boldsymbol  B_{n,R_x}^{v}]_{1,1})^{T}$,  $\boldsymbol{s} = (s_1,s_2,...,s_{R_x})^{T}$. 
Since the denominators of $(\ref{deno1})-(\ref{deno4})$ are still quadratic functions of $\{s_r\}$ and $(\boldsymbol {Q}_n^{d}$, $\boldsymbol {Q}_n^{v})$ are neither positive definite nor negative definite matrices, it makes the problem $(\ref{B})$ hard to obtain global optimum at low complexity. To address it, we prove that $(\ref{B}a)$ and $(\ref{B}b)$ can be converted into convex constraints by following steps
\begin{align}
	 \frac{\boldsymbol{s}^{T}\boldsymbol{p}}{\boldsymbol{s}^{T}\boldsymbol{Q}\boldsymbol{s}} \leq \eta\\
\Longleftrightarrow-\eta \boldsymbol{s}^{T}\boldsymbol{Q}\boldsymbol{s} + \boldsymbol{s}^{T}\boldsymbol{p} \leq 0\\
 \Longleftrightarrow-\boldsymbol{s}^{T}(\boldsymbol{Q}+\boldsymbol{Q}^T)\boldsymbol{s} + \frac{2}{\eta}\boldsymbol{s}^{T}\boldsymbol{p} \leq 0 \label{eq}.
\end{align}
We denote $\boldsymbol{Z} = -(\boldsymbol{Q}+\boldsymbol{Q}^T)$ and $\lambda_{\boldsymbol{Z}} $ as the minimal eigenvalue of $\boldsymbol{Z}$. Then, $(\ref{eq})$ is equivalently to
\begin{align}
	\boldsymbol{s}^{T}(\boldsymbol{Z}-\lambda_{\boldsymbol{Z}}\boldsymbol I)\boldsymbol{s} + \frac{2}{\eta}\boldsymbol{s}^{T}\boldsymbol{p}+ \lambda_{\boldsymbol{Z}}\boldsymbol{s}^{T}\boldsymbol I\boldsymbol{s}\leq 0 \label{psd}.
\end{align}
Since $\boldsymbol{s}^{T}\boldsymbol I\boldsymbol{s} = \sum\limits_{r=1}^{R_x}s_r= N_r$ and $\boldsymbol{Z}-\lambda_{\boldsymbol{Z}}\boldsymbol I$ is positive semidefinite, $(\ref{psd})$ is a convex constraint. Therefore, for given $\eta_v$, $(\ref{B})$ can be equivalently converted to the following convex quadratic integer problem which can be efficiently solved by optimization tools such as $Mosek$,
\begin{align}
	 \min&_{\{s_r, \eta_d\}}\quad \eta_d  \label{B'} \\
	\emph{s.t.}:\quad
	& 	\boldsymbol{s}^{T}(\boldsymbol{Z}_{n}^{d}-\lambda_{\boldsymbol{Z}_{n}^{d}}\boldsymbol I)\boldsymbol{s} + \frac{2}{\eta_d}\boldsymbol{s}^{T}\boldsymbol{p}_{n}^{d,1}+ \lambda_{\boldsymbol{Z}_{n}^{d}}N_r\leq 0 \tag{\ref{B'}a}\\
	& 	\boldsymbol{s}^{T}(\boldsymbol{Z}_{n}^{d}-\lambda_{\boldsymbol{Z}_{n}^{d}}\boldsymbol I)\boldsymbol{s} + \frac{2}{\eta_d}\boldsymbol{s}^{T}\boldsymbol{p}_{n}^{d,2}+ \lambda_{\boldsymbol{Z}_{n}^{d}}N_r\leq 0 \tag{\ref{B'}b}\\
	& 	\boldsymbol{s}^{T}(\boldsymbol{Z}_{n}^{v}-\lambda_{\boldsymbol{Z}_{n}^{v}}\boldsymbol I)\boldsymbol{s} + \frac{2}{\eta_v}\boldsymbol{s}^{T}\boldsymbol{p}_{n}^{v,1}+ \lambda_{\boldsymbol{Z}_{n}^{v}}N_r\leq 0 \tag{\ref{B'}c}\\
	& 	\boldsymbol{s}^{T}(\boldsymbol{Z}_{n}^{v}-\lambda_{\boldsymbol{Z}_{n}^{v}}\boldsymbol I)\boldsymbol{s} + \frac{2}{\eta_v}\boldsymbol{s}^{T}\boldsymbol{p}_{n}^{v,2}+ \lambda_{\boldsymbol{Z}_{n}^{v}}N_r\leq 0 \tag{\ref{B'}d}\\
	& \sum_{r=1}^{R_x}s_r = N_r \tag{\ref{B'}e}
\end{align}
Where $\boldsymbol{Z}_{n}^{d} = -( \boldsymbol{Q}_n^d+(\boldsymbol{Q}_n^d)^T )$ and $\boldsymbol{Z}_{n}^{v} = -( \boldsymbol{Q}_v^d+(\boldsymbol{Q}_v^d)^T )$.
The problem $(\ref{B'})$ is essentially a feasible problem and bisection search method is applied for searching for the optimal $\eta_d$. The details of solving $(\ref{B'})$ is described in Algorithm 2.
\begin{algorithm}
    \caption{Bisection Search Method for Solving Problem $(\ref{B})$}
    \label{alg:2}
    \begin{algorithmic}
        \REQUIRE $\eta_l^d $, $\eta_h^d$, tolerance $\epsilon$;\\Make sure $(\ref{B'})$ is feasible for $\eta_h^d$ and infeasible for $\eta_l^d$.\
        \REPEAT
        \STATE $\eta^d$ = $\frac{\eta_l^d+\eta_h^d}{2}$;
        \STATE solve $(\ref{B'})$;
            \IF{$\{'feasible'\}$}
        \STATE $\eta_h^d$ = $\eta^d$;
    			\ELSE
        \STATE $\eta_l^d$ = $\eta^d$;
    			\ENDIF
        \UNTIL{$\{\eta_h^d-\eta_l^d\leq\epsilon\}$}
		\ENSURE $\eta^d$.
    \end{algorithmic}
\end{algorithm}

\begin{figure*}[h]
	\centering  
	\subfloat[different distributions of radar receivers]{
		\includegraphics[width=5.85cm,height=5.85cm]{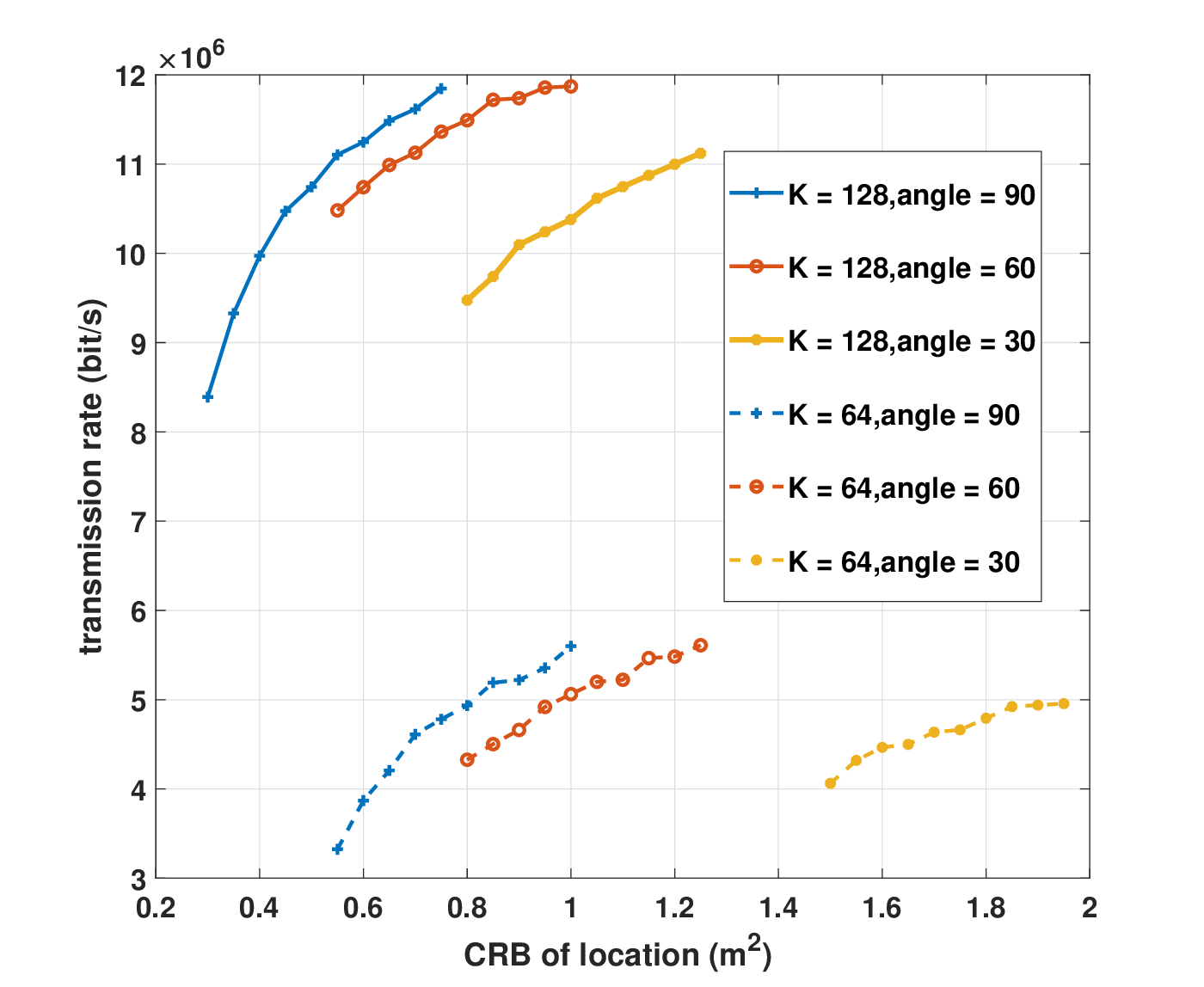}}
	\subfloat[different numbers of radar receivers]{
		\includegraphics[width=5.85cm,height=5.85cm]{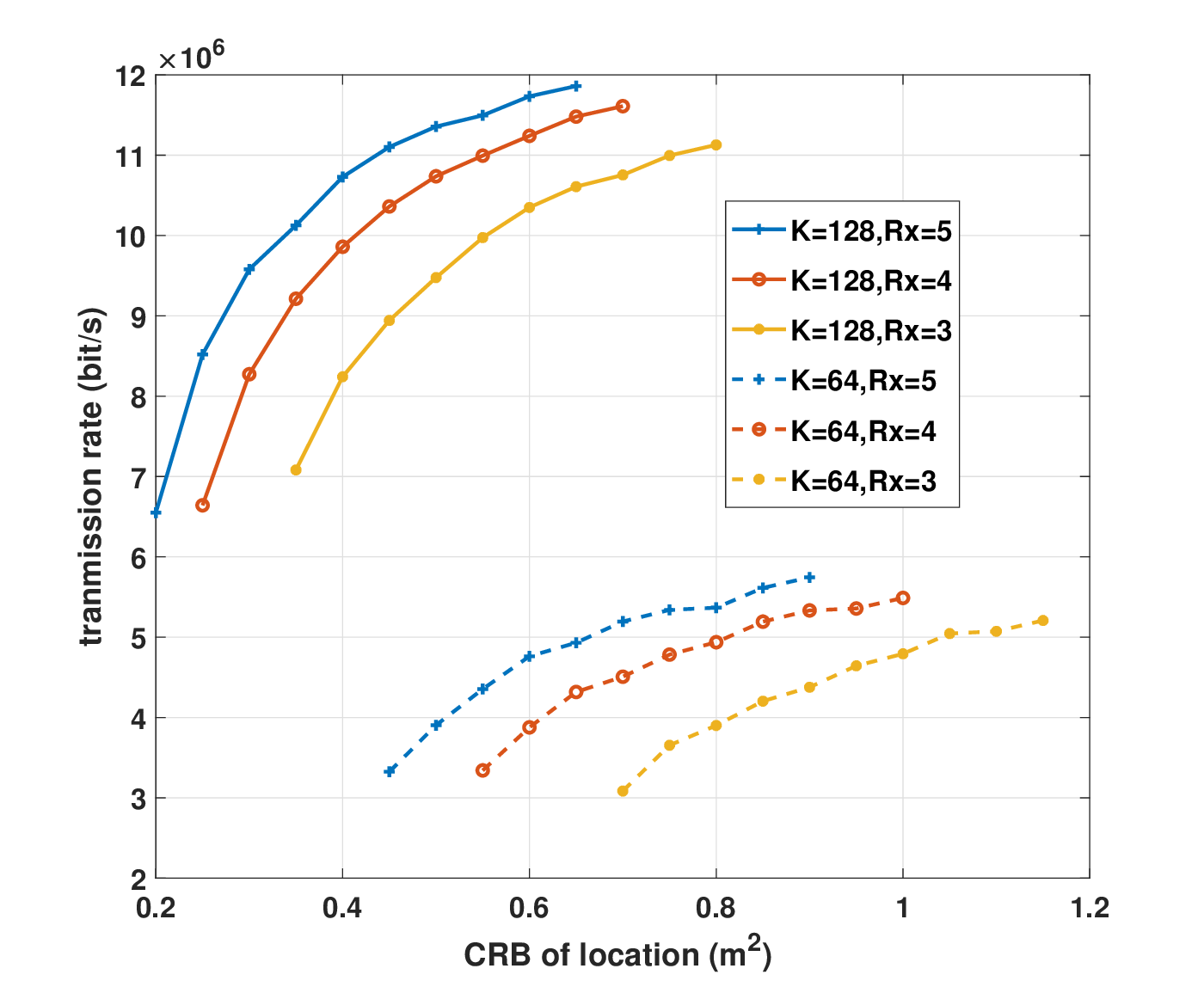}}
	\subfloat[different numbers of detection subareas]{
		\includegraphics[width=5.85cm,height=5.85cm]{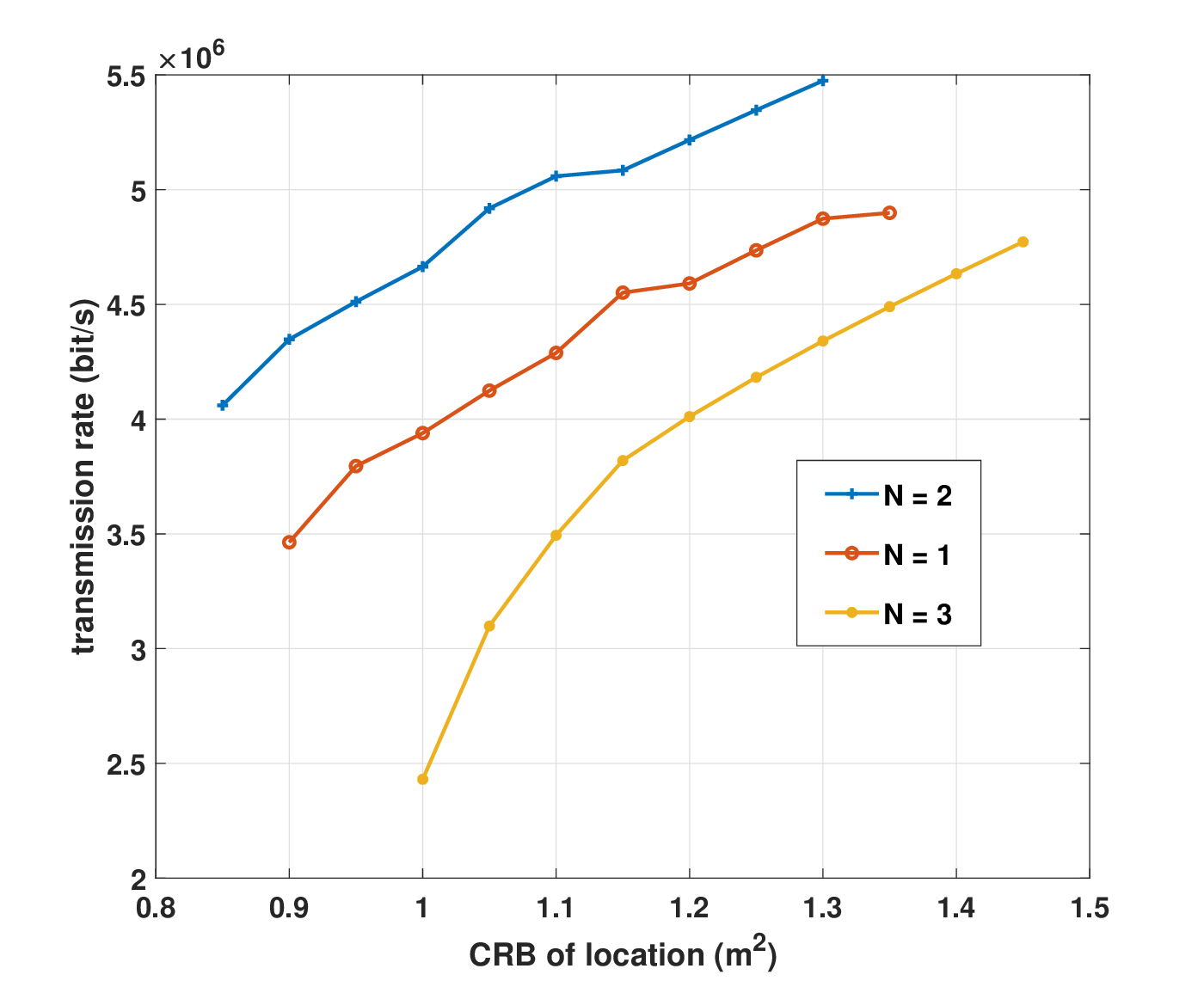}}
	\caption{The tradeoff between transmission rate and  CRB of location with different parameters configurations.}
    \label{fig_location}
\end{figure*}
\begin{figure*}[h]
	\centering  
	\subfloat[different distributions of radar receivers]{
		\includegraphics[width=5.85cm,height=5.85cm]{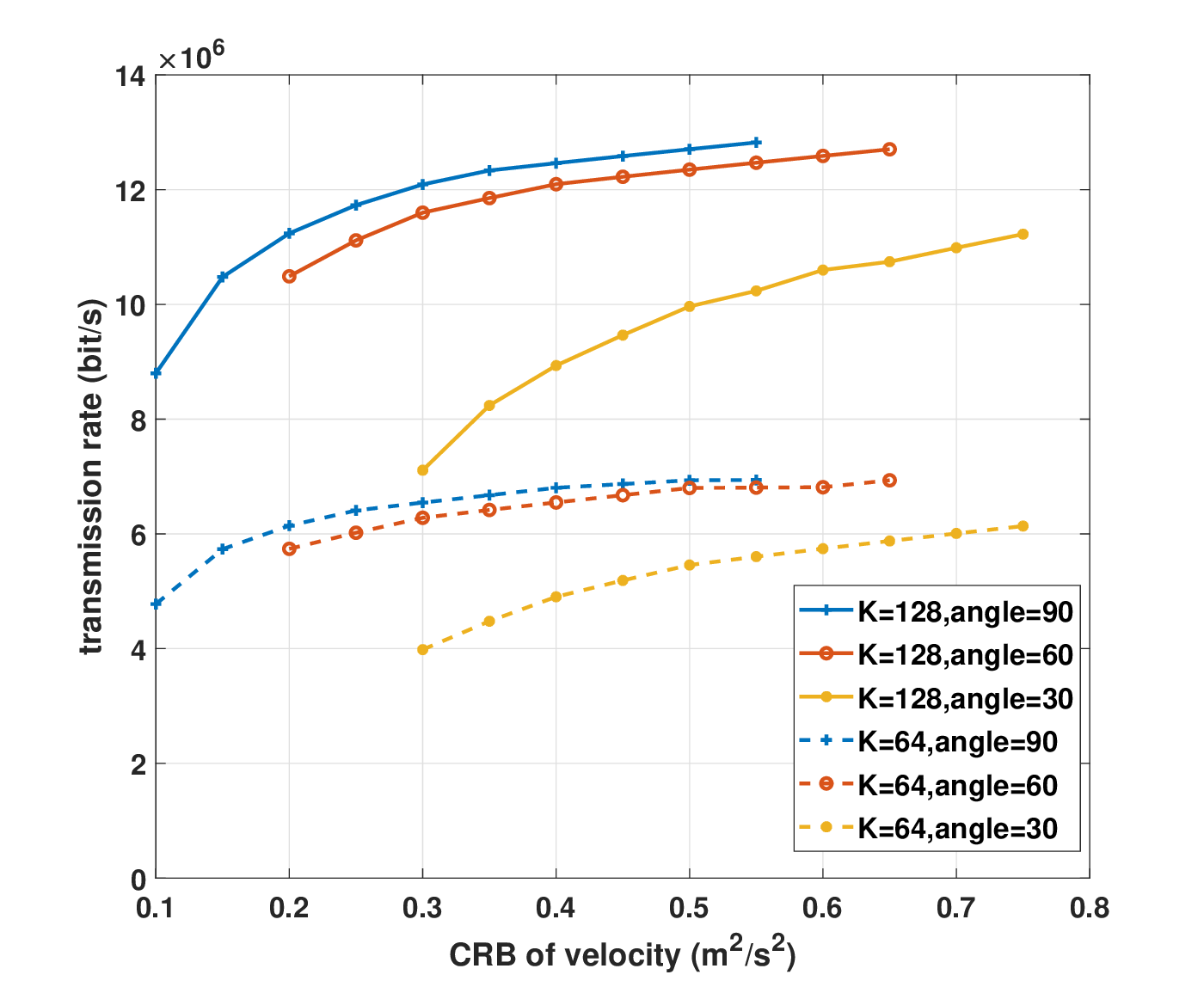}}
	\subfloat[different numbers of radar receivers]{
		\includegraphics[width=5.85cm,height=5.85cm]{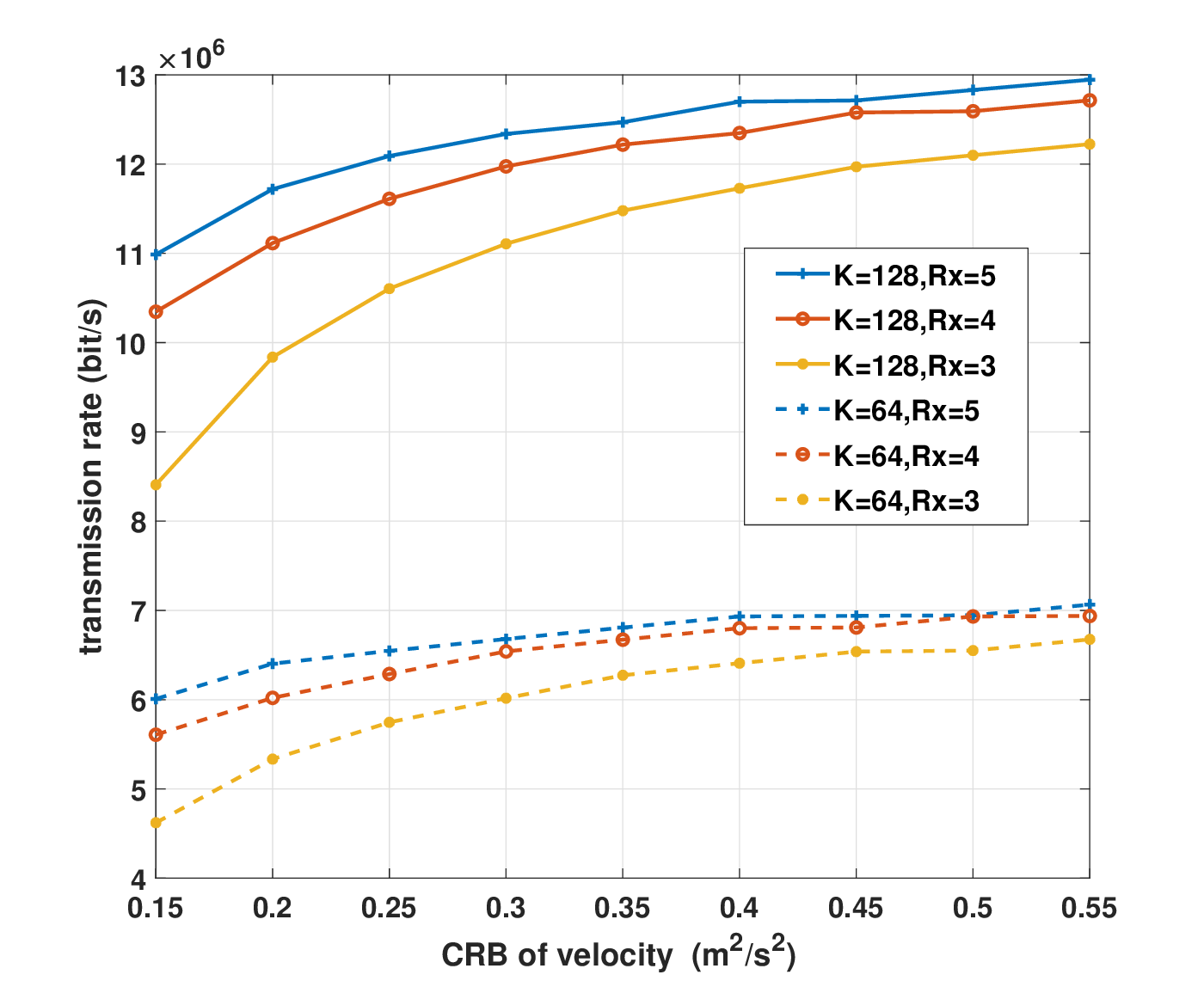}}
	\subfloat[different numbers of detection subareas]{
		\includegraphics[width=5.85cm,height=5.85cm]{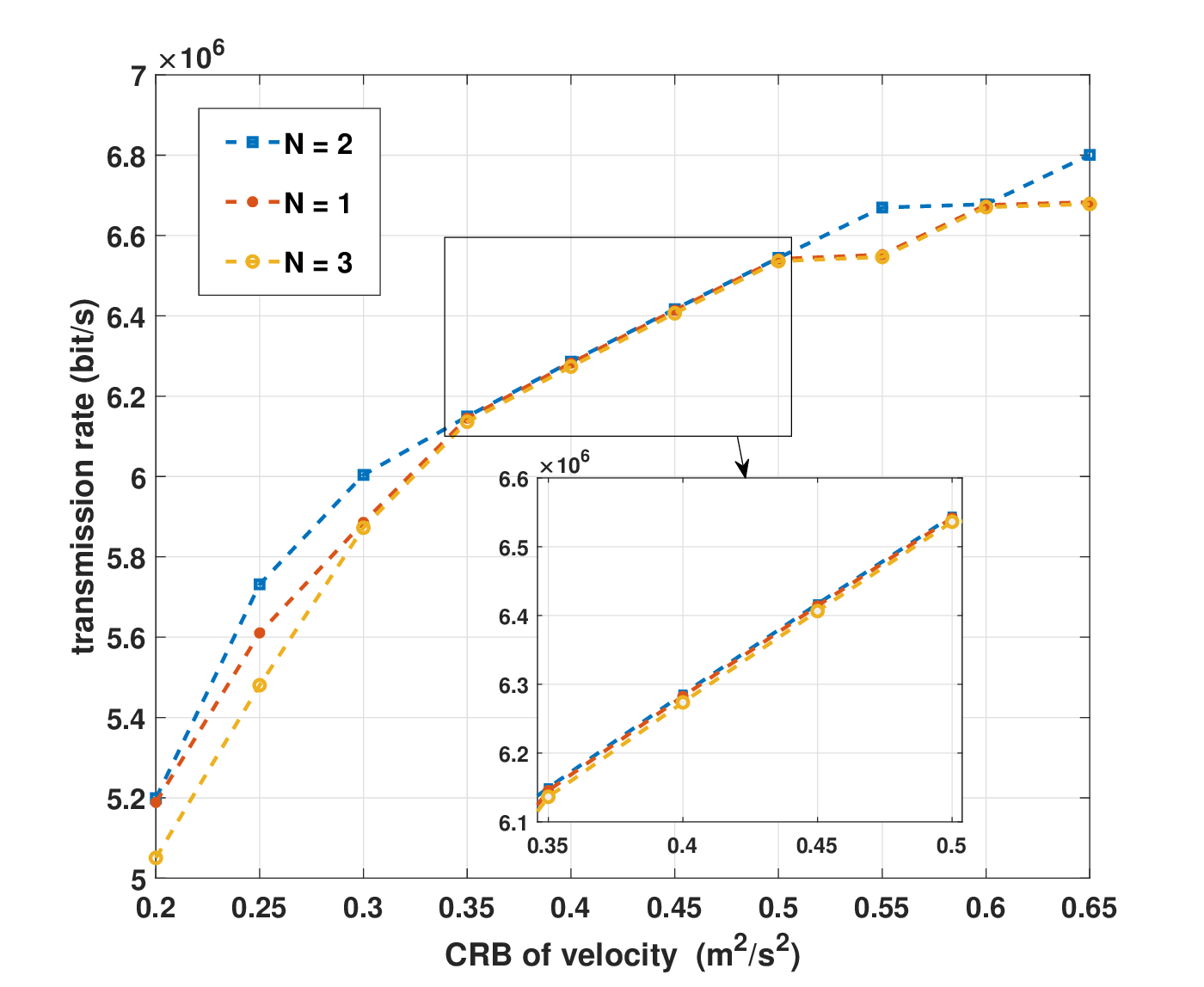}}
	\caption{The tradeoff between  transmission rate and CRB of velocity with different parameters configurations.}
	\label{fig_velocity}
\end{figure*}


\section{Numerical Results}
In this section, we provide simulation results to show the the performance of distributed MIMO-OFDM DFRC system with different parameters settings. We set $f_c= 3$ GHz, $T_x = 32$, $\Delta f = 15$KHz, $T_{cp} = 4.7\mu s$, $T = 71.7\mu s$, $P_{max} = 5W$, $L = 32$. The noise power of radar receivers and communication receiver are $\sigma_{\bar{w}}^2 =1.5\times 10^{-18}W$ and $ \sigma_z^2 = 1.5\times 10^{-14}W$, respectively. The attenuation coefficients in $(\ref{CRB3})$ are set to $c_{n,r} = \sqrt{ \frac{\lambda^2 RCS_r}{(4\pi)^3(d_{n,0})^2(d_{n,r})^2} }$ where $RCS_r$ is Radar Cross Section to the $r$-th radar receiver, $d_{n,0}$ is the distance from transmit BS to the $n$-th target and $d_{n,r}$ is the distance from the $n$-th target to the $r$-th radar receiver. We assume $RCS_r$ is uniformly distributed between 0.09 to 0.1 \cite{102}. The propagation loss $a_m$ in $(\ref{noise_c})$ is set to $a_m= \sqrt{ \frac{\lambda^2}{(4\pi)^2 d_{m}^2} }$ where $d_m$ represents the distance between the BS and the $m$-th communication user.
\begin{figure}[h]
\centering
\includegraphics[width=7.5cm,height=6.5cm] {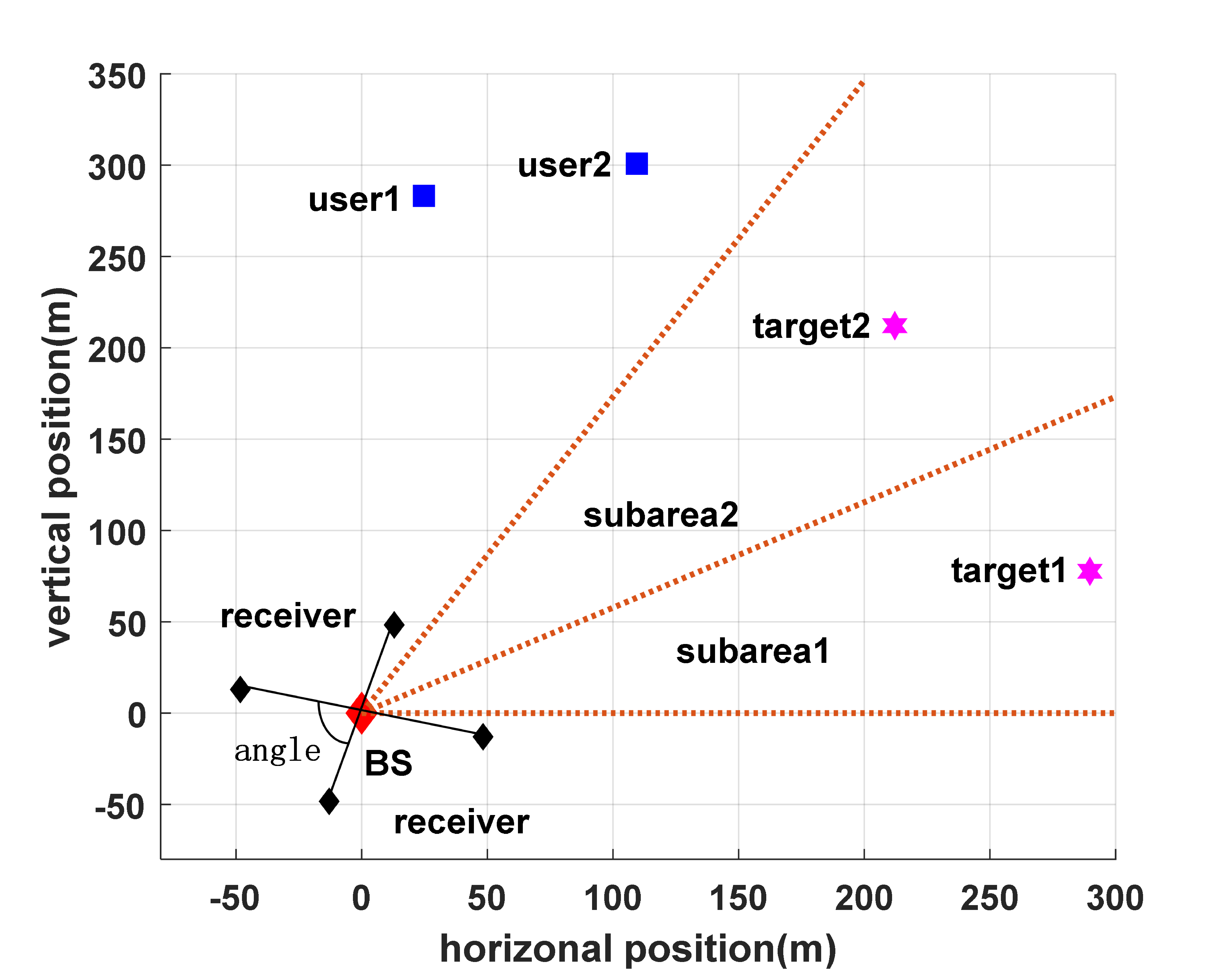}
\caption{The position of the BS, receivers, targets and users.}
\label{MIMO OFDM system}
\end{figure}
\begin{figure}[h]
\centering
\includegraphics[width=7.5cm,height=6.5cm] {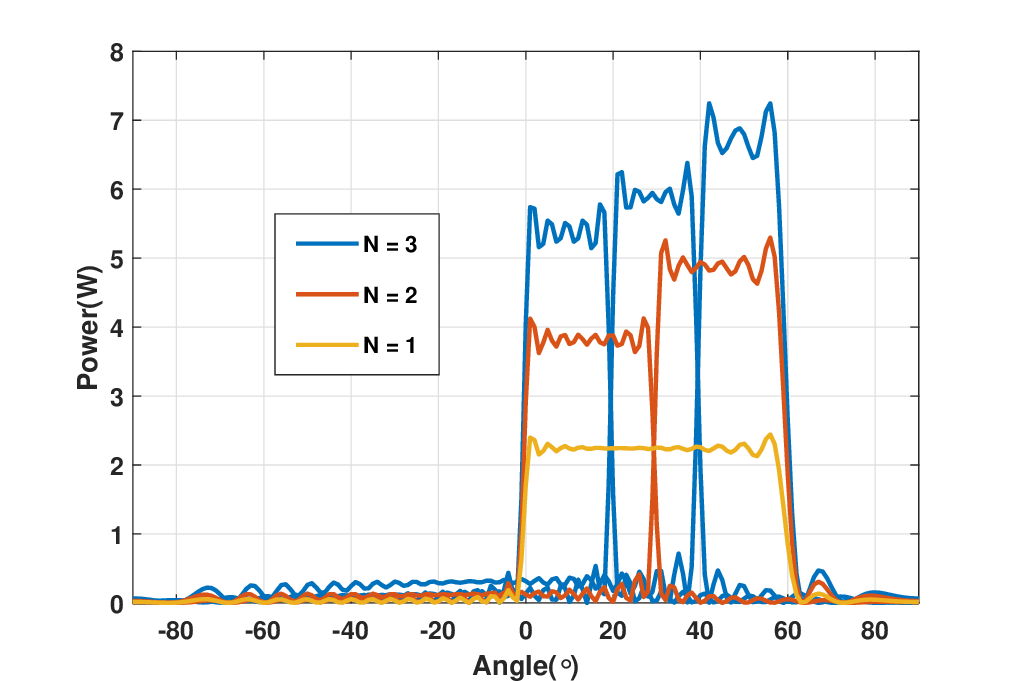}
\caption{Beampatterns under different number of detection subareas.}
\label{beamwithdiffN}
\end{figure}
\subsection{Tradeoff between Communication and sensing Performance with Subcarrier/Power Allocation}
In this subsection, we demonstrate the effects of different parameters on the tradeoff between communication and radar performance by solving $(\ref{A})$ with Algorithm 1. Specifically, given estimation accuracy constraints $\eta_d$ and $\eta_v$ and other parameters, we solve $(\ref{A})$ to obtain the subcarrier/power allocation scheme. Then, we can calculate the transmission rate $\frac{1}{T}\sum_{k=1}^{K}\sum_{m=1}^{M}R_{k,m}$ and the CRB of location and velocity according to $(\ref{Cd})$ and $(\ref{Cv})$, respectively. By varying $\eta_d$ or $\eta_v$, tradeoffs between transmission rate and CRB are exhibited in Fig.2 and Fig.3.

In Fig.$\ref{fig_location}$(a) and Fig.$\ref{fig_velocity}$(a), we demonstrate the relationship between transmission rate and CRB of location and velocity with three different radar receivers placements. We set $M = 2$, $N = 2$, $R_x=4$ and $K = 64,128$, the relative positions of BS, communication users, targets and radar receivers are showed in Fig.$\ref{MIMO OFDM system}$. The location of BS is $[0m,0m]$ and the radar receivers are distributed on a circle with the BS as the center and 50 meters as the radius. The value of  `angle' in Fig.$\ref{MIMO OFDM system}$ represents the angular spacing between radar receivers. Specifically, `angle = 90' means that the four radar receivers are located at angles $-15^\circ,75^\circ,165^\circ,255^\circ$ respectively, `angle = 60' means that the four radar receivers are located at angles $0^\circ,60^\circ,180^\circ,240^\circ$ respectively,  `angle = 30' means that the four receivers are located at angles $15^\circ,45^\circ,195^\circ,225^\circ$ respectively. The angel range of the detection area $[0^\circ,60^\circ]$ is divided into two subareas and each subarea has an angular coverage of $30^\circ$.
We assume that there are two targets located within the detection area, with a velocity of 20m/s and locations of $[289.8m,77.6m]$ and $[212.1m,212.1m]$.
The locations of communication users are  $[24.8m,283.2m]$ and $[109.5m,300.8m]$ respectively.
It is observed that reducing the detection performance results in more resources being allocated to communication users, thus increasing the communication rate. The right most points of the curves show the CRB of the targets in the detection area when all resources are allocated to communication users. We can see that when the locations of the communication users are near the detection area, a good detection performance can still be achieved even if all the subcarriers are allocated to communication users. Increasing the number of subcarriers (signal bandwidth) can effectively improve the location estimation performance, but the improvement of the velocity estimation performance is not significant.
\begin{figure}[h]
\begin{minipage}[t]{0.47\textwidth}
\includegraphics[width=\textwidth] {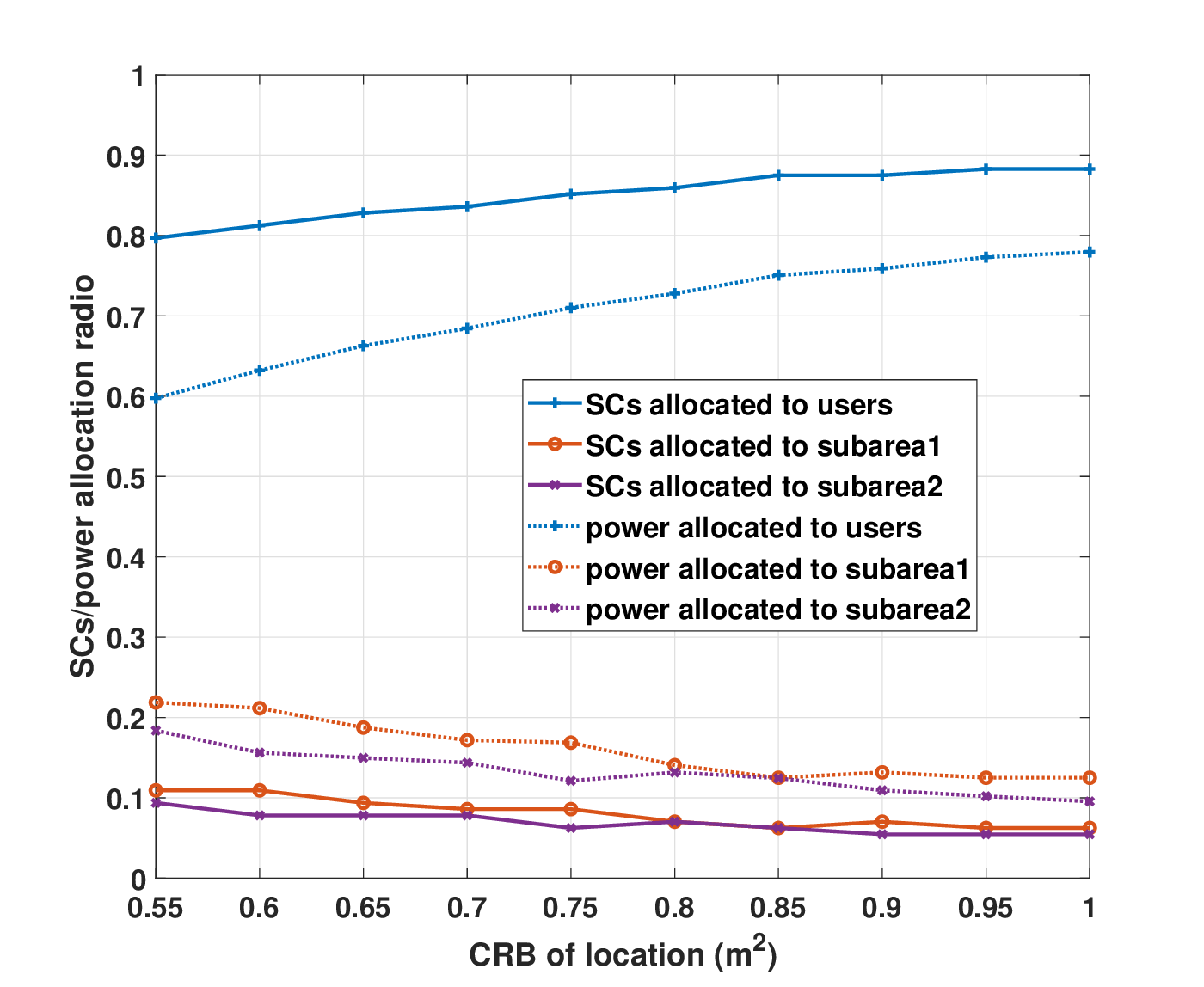}
\caption{Subcarrier and power allocation radio versus CRB of location.}
\label{SC_power_radio_location}
\end{minipage}
\begin{minipage}[t]{0.47\textwidth}
\includegraphics[width=\textwidth] {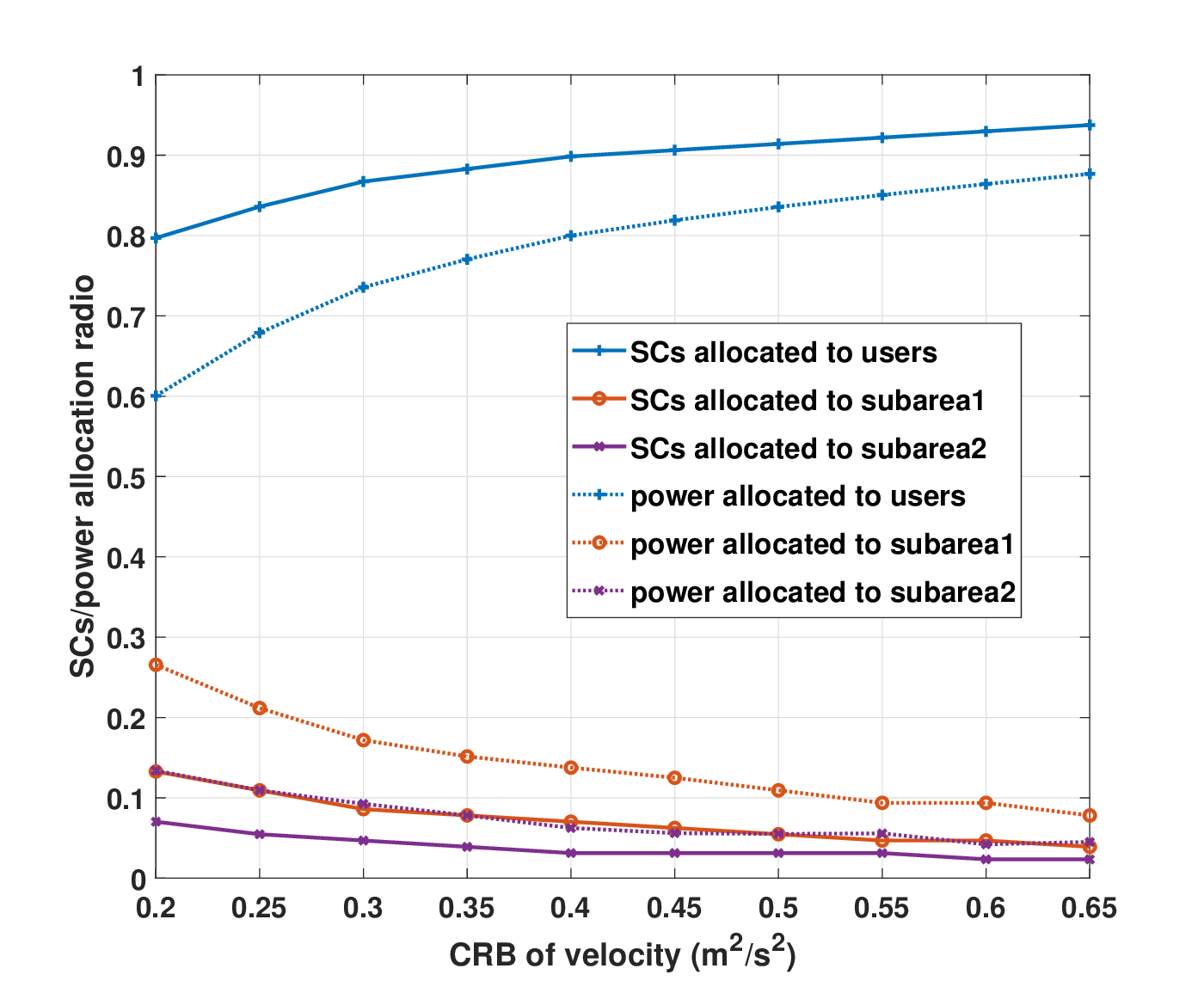}
\caption{Subcarrier and power allocation radio versus CRB of velocity.}
\label{SC_power_radio_velocity}
\end{minipage}
\end{figure}
When the $RCS_r$ of different radar receivers are approximately equal, the more evenly distributed the receivers are, the better the estimation performance is.
The estimation performance is optimal when the minimum distance between radar receivers is maximized.

\begin{figure*}[h]
\begin{minipage}[t]{0.5\textwidth}
\includegraphics[width=\textwidth] {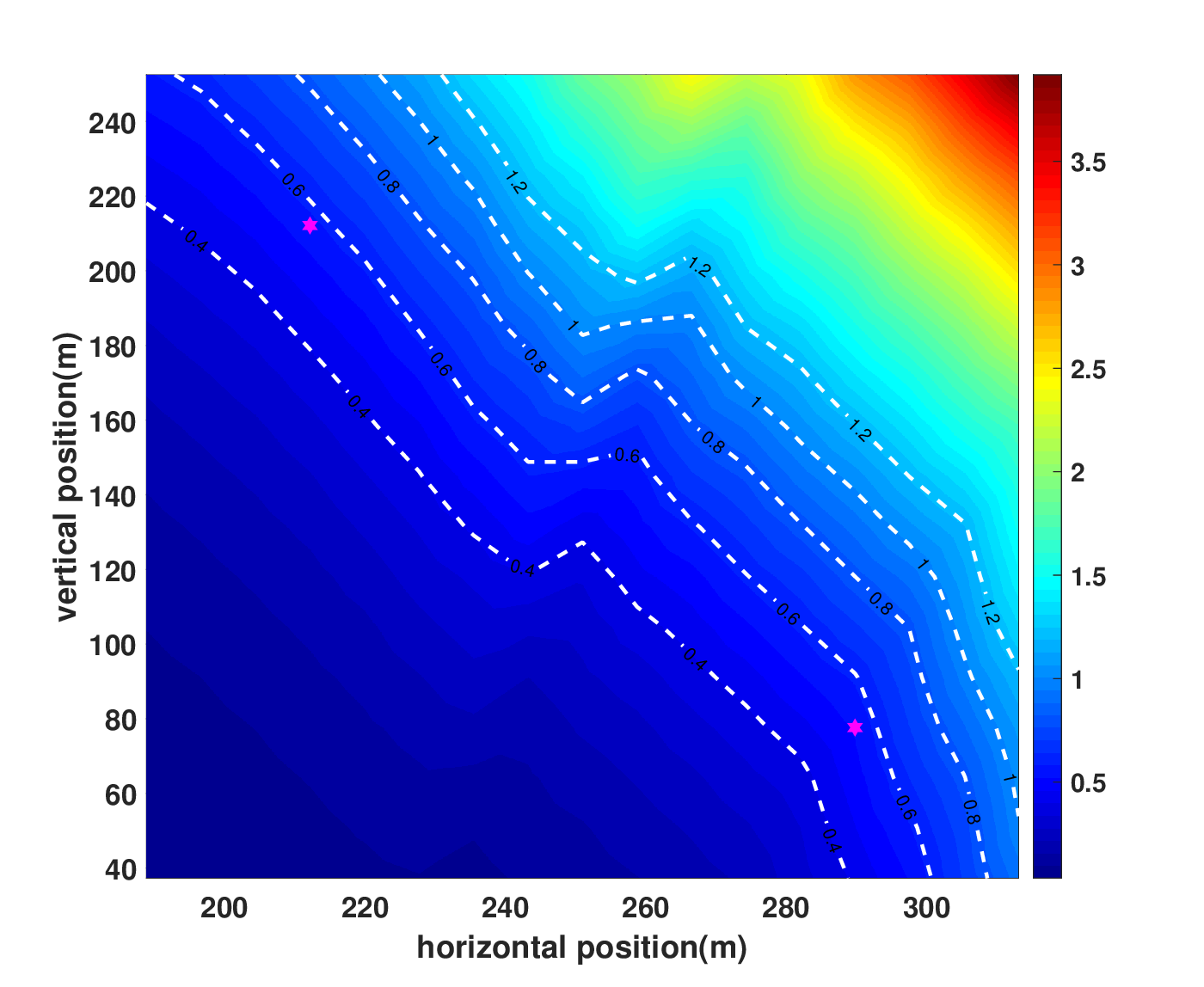}
\caption{Spatial distribution of CRB of location near the targets.}
\label{around_CRB_location}
\end{minipage}
\begin{minipage}[t]{0.5\textwidth}
\includegraphics[width=\textwidth] {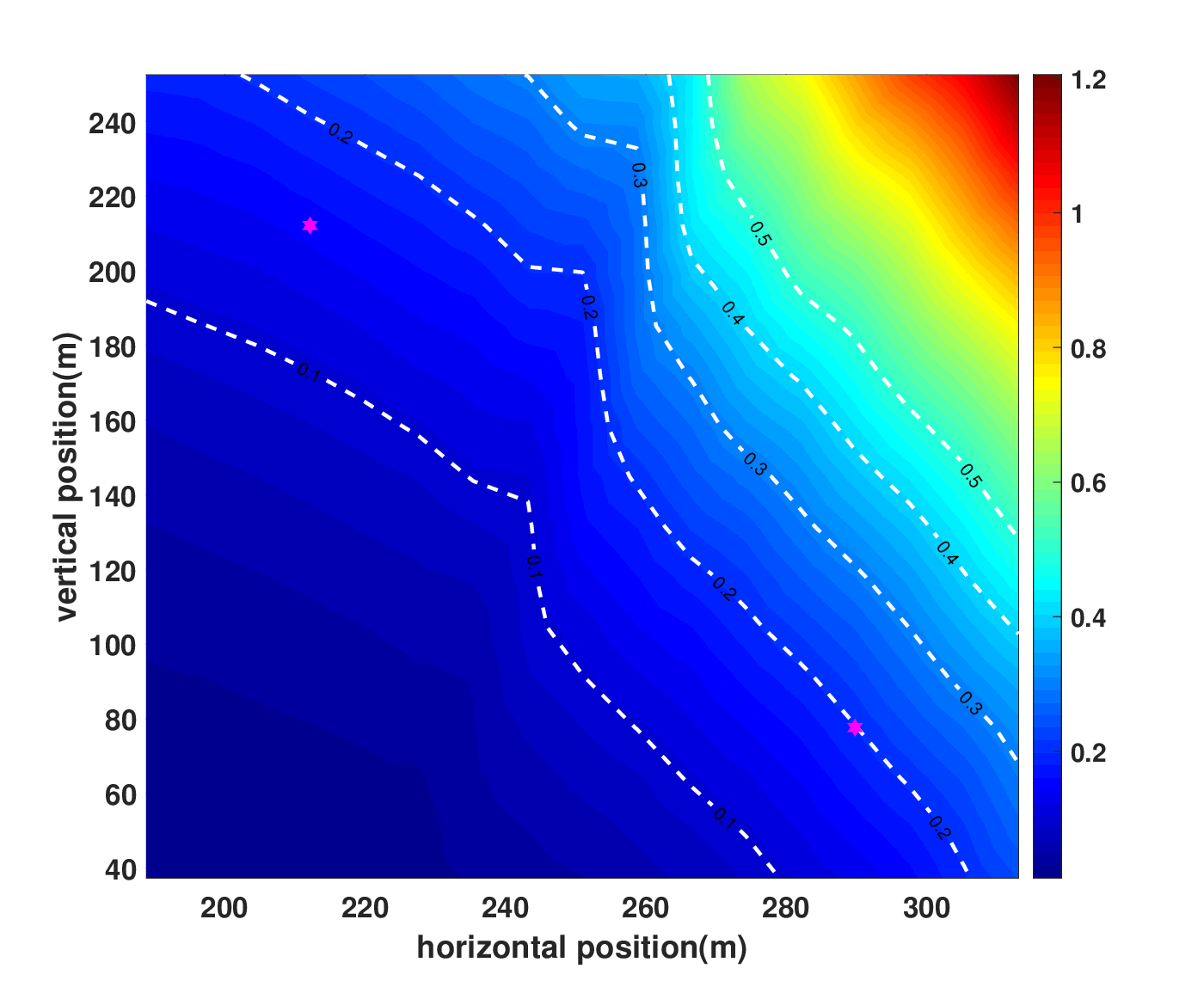}
\caption{Spatial distribution of CRB of velocity near the targets.}
\label{around_CRB_velocity}
\end{minipage}
\end{figure*}

Fig.$\ref{fig_location}$(b) and Fig.$\ref{fig_velocity}$(b) show the relationship between transmission rate and CRB of location and velocity estimation with different numbers of radar receivers. All parameter settings are the same as Fig.$\ref{fig_location}$(a) and Fig.$\ref{fig_velocity}$(a) except for the number of radar receivers. $R_x = 3$ means the three radar receivers are located at angles $0^\circ,120^\circ,240^\circ$ respectively, $R_x = 4$ means that the four radar receivers are located at angles $0^\circ,90^\circ,180^\circ,270^\circ$ respectively and $R_x = 5 $ means that the five radar receivers are located at angles $0^\circ,72^\circ,144^\circ,216^\circ,288^\circ$ respectively.
Obviously, increasing the number of cooperative receivers will improve the estimation performance. However, The diversity gain brought by distributed receivers gradually decreases as the number of receivers increases.
Moreover, Fig.$\ref{fig_location}$(a) and Fig.$\ref{fig_location}$(b) show that both higher transmission rate and lower CRB of location can be achieved when $K$ increases from 64 to 128. However, Fig.$\ref{fig_location}$(a) and Fig.$\ref{fig_velocity}$(b) show that, as $K$ increases, the CRB of velocity estimation is not affected and only transmission rate is improved. The reason is that the elements of $\boldsymbol{D}_{n}(r,k,n')$ are positively correlated with $K$ according to $(\ref{D11})-(\ref{D22})$, while the elements of $\boldsymbol{V}_{n}(r,k,n')$ are independent of $K$ according to $(\ref{V11})-(\ref{V22})$.
\begin{figure*}[h]
\begin{minipage}[t]{0.5\textwidth}
\includegraphics[width=\textwidth] {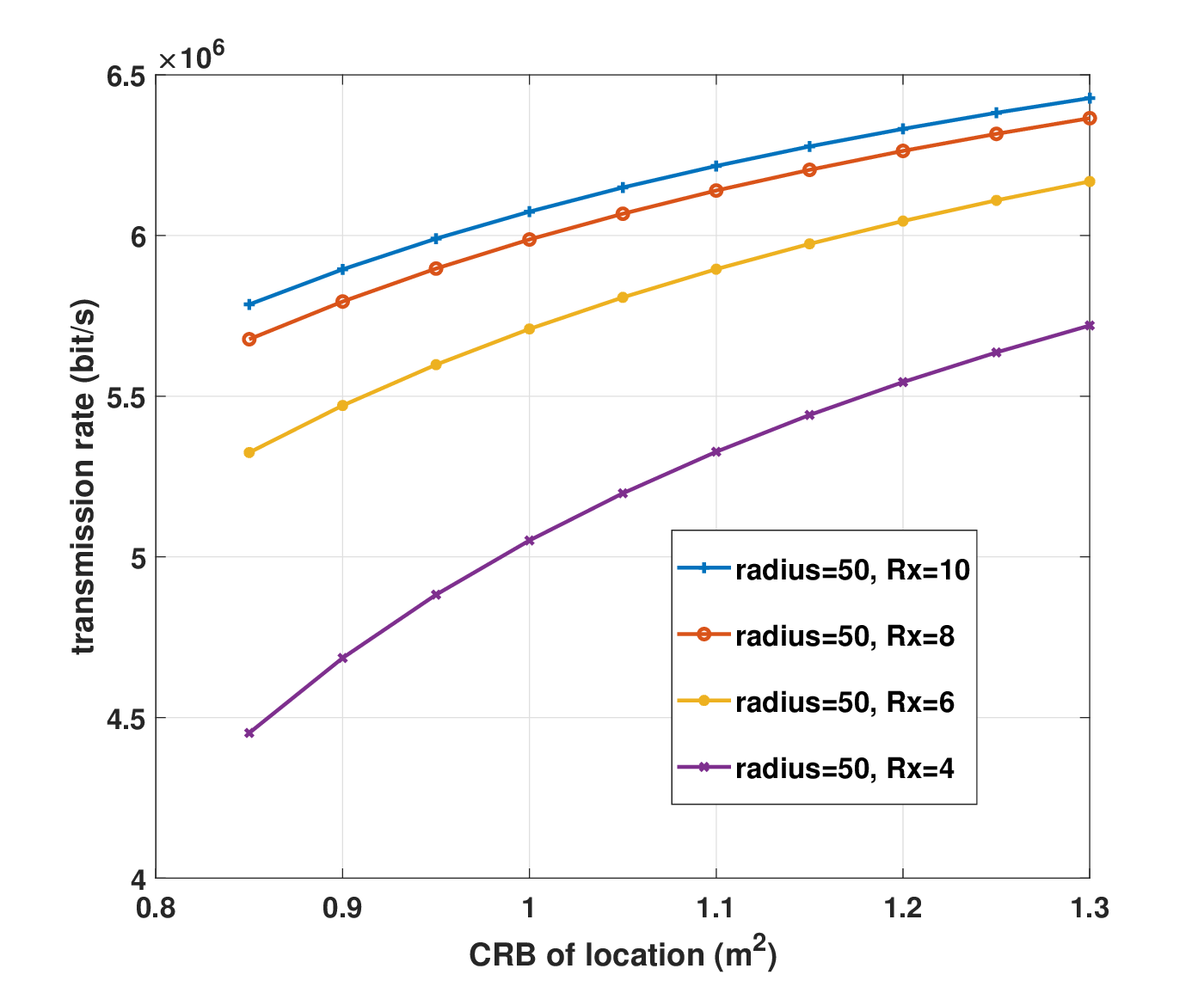}
\caption{Transmission rate vs. the CRB of location with different $R_x$.}
\label{}
\end{minipage}
\begin{minipage}[t]{0.5\textwidth}
\includegraphics[width=\textwidth] {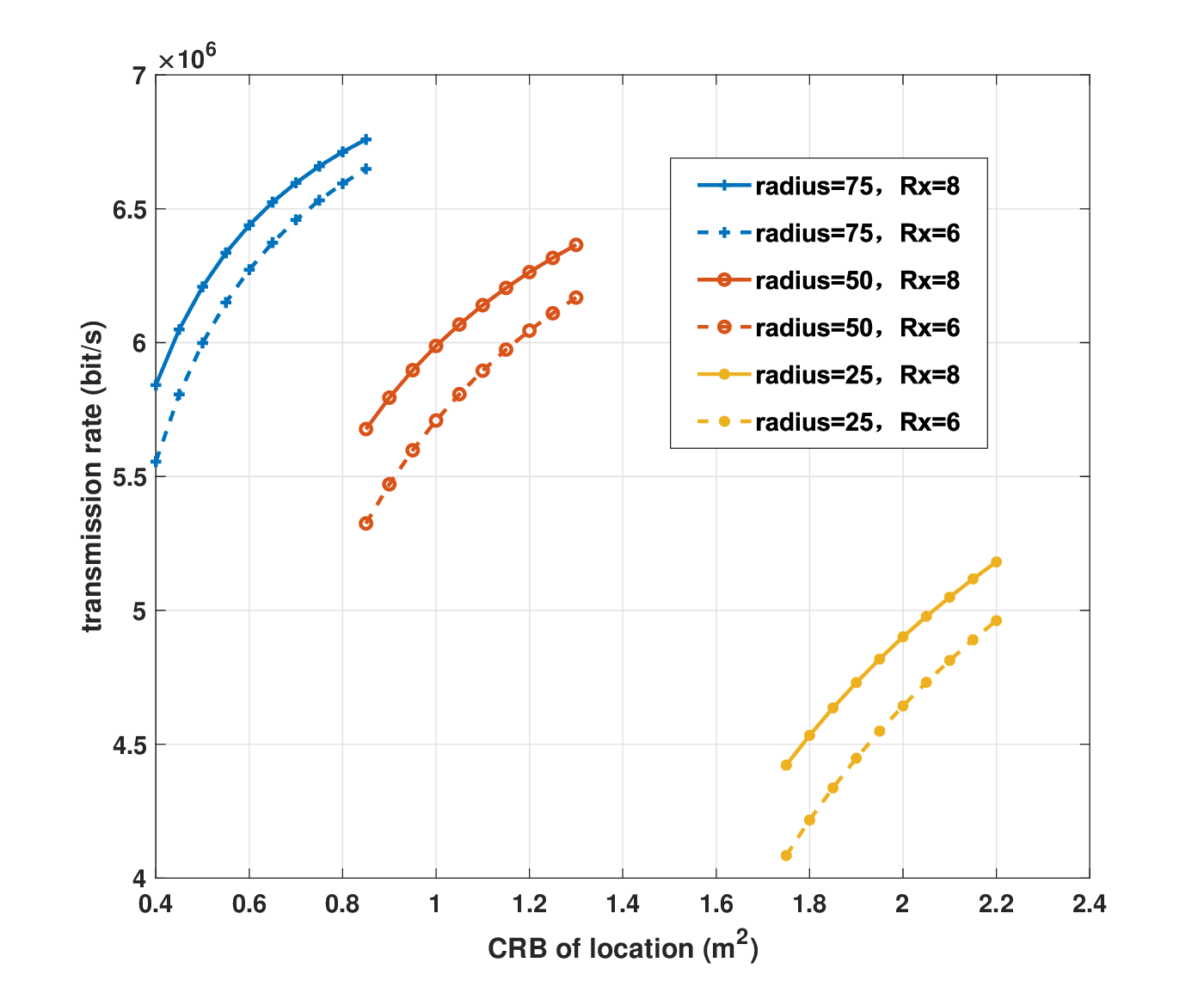}
\caption{Transmission rate vs. the CRB of location with different radius.}
\label{}
\end{minipage}
\end{figure*}
\begin{figure*}[h]
\begin{minipage}[t]{0.5\textwidth}
\includegraphics[width=\textwidth] {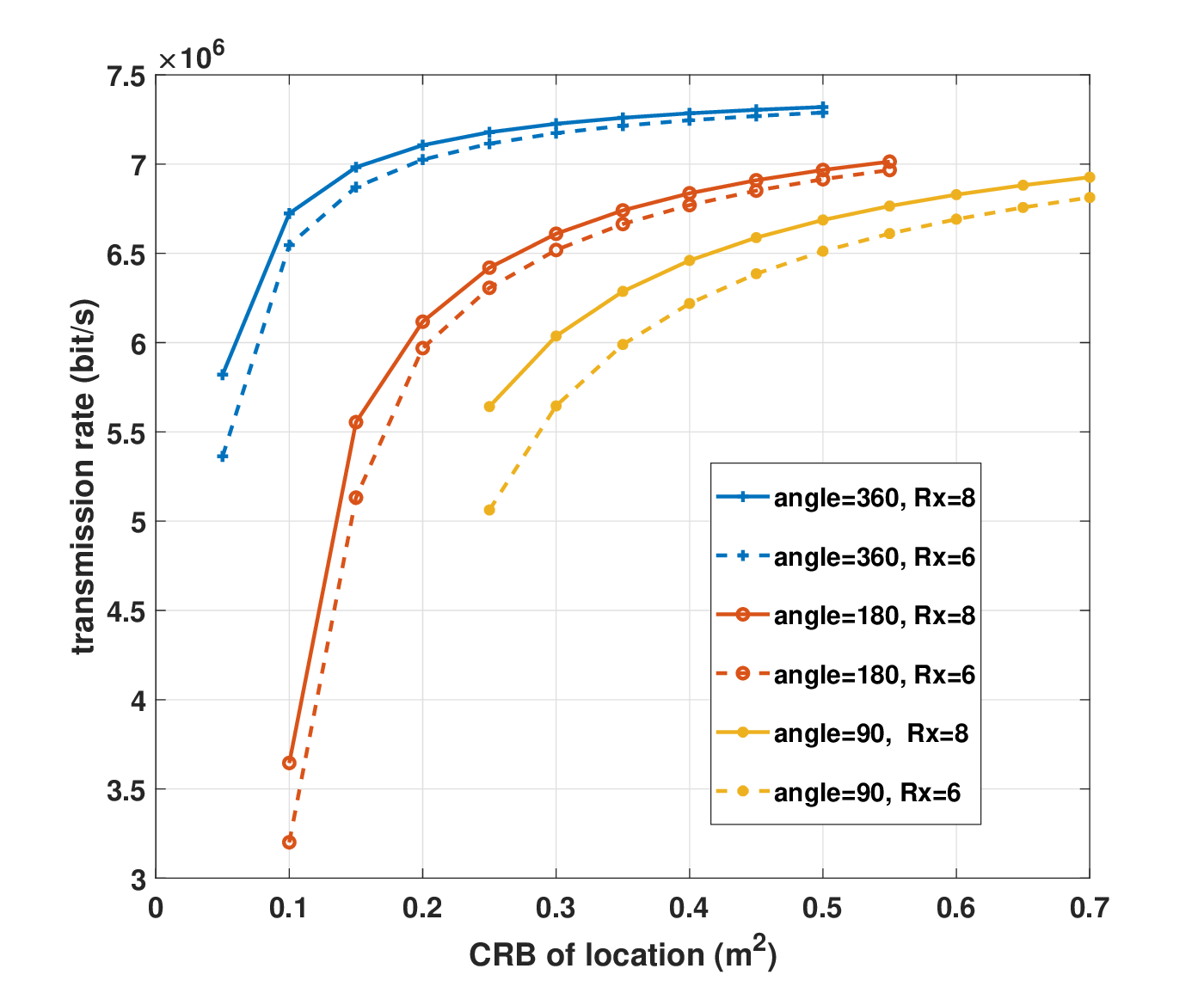}
\caption{Transmission rate vs. the CRB of location with different angles.}
\label{target_location1}
\end{minipage}
\begin{minipage}[t]{0.5\textwidth}
\includegraphics[width=\textwidth] {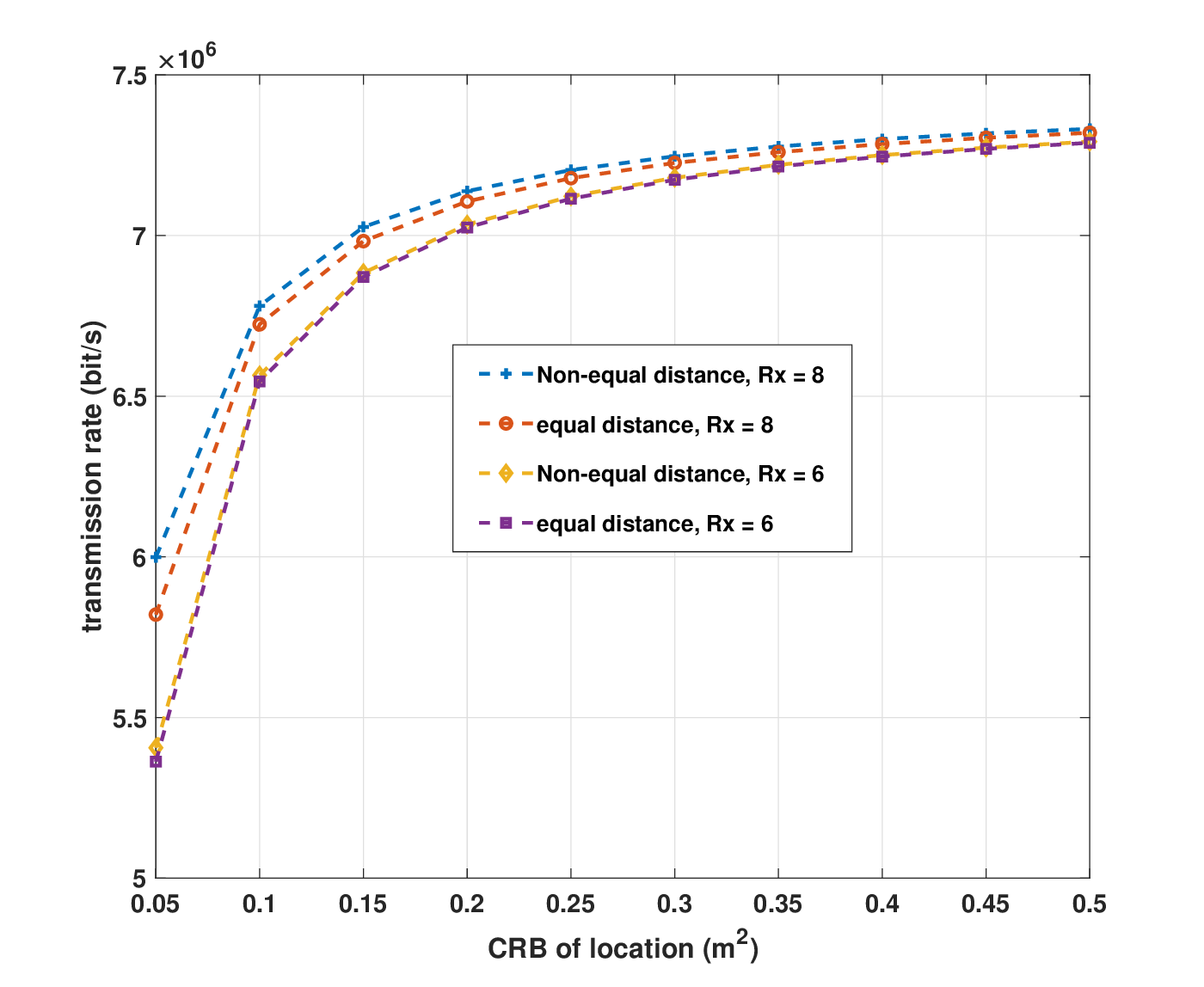}
\caption{Transmission rate vs. the CRB of location with angle and $R_x$.}
\label{target_location2}
\end{minipage}
\end{figure*}

Fig.$\ref{fig_location}$(c) and Fig.$\ref{fig_velocity}$(c) show the tradeoff between transmission rate and CRB of location and velocity estimation with different numbers of detection subareas. The detection beampatterns will change with the varying of $N$. Taking $N=2$ as an example, the angular range $[0^\circ,60^\circ]$ of the whole detection area is divided equally into two detection subareas $[0^\circ,30^\circ],[30^\circ,60^\circ]$ as shown in Fig.4. Different subcarrier sets are used to probe different subareas. The curves of $N=2$ in Fig. 5 show the beampattern when two sets of subcarriers are used to probe two subareas separately. The beampatterns with $N=1$ and $N=3$ are also plotted for comparison. From Fig.$\ref{fig_location}$(c) and Fig.$\ref{fig_velocity}$(c) we can see that the tradeoff performance in the case of $N = 2$ is better than the cases of $ N = 1$ and $N = 3$. The performance of $N= 2$ is better than that of $N= 3$ because as $N$ increases, more subcarriers are used to detect, which reduces the transmission rate. On the other hand,   the power gain of beampatterns of $N = 2$ is significantly higher than that of $N=1$, which helps to use fewer subcarriers for detection and thus increases the transmission rate.

In Fig.$\ref{SC_power_radio_location}$ and Fig.$\ref{SC_power_radio_velocity}$, subcarriers and power allocation schemes are presented. We set $N=2$, $R_x = 4$, and $K = 128$, and the placement of radar receivers is the same as the case of ``angle = 60" in Fig.{\ref{MIMO OFDM system}}. It can be seen that as CRBs increase, the proportion of subcarriers and power allocated to the communication users gradually increases. In addition, the proportion of subcarriers and power allocated to target in the subarea 1 is always lower than that allocated to target in the subarea 2, because the beam power gain towards the subarea 1 is smaller than that toward the subarea 2. Therefore, target 1 needs more subcarriers and power to achieve the same detection accuracy as target 2.

Fig.\ref{around_CRB_location} and Fig.\ref{around_CRB_velocity} show the spatial distribution of CRBs near detection targets. The red stars represent the locations of detection targets. We set $N=2$, $R_x = 4$, $K = 128$, $\eta_d = 0.55$, $\eta_v = 0.2$, and the distribution of radar receivers is the same as the case of $angle = 60^\circ$ in Fig.4. By solving problem (26) we can obtain the solution of subcarrier
and power allocation, based on which we calculate the CRBs of location and velocity of the sampled points near the targets. The CRB contours outline the detection area in which distance estimation error is less than 1.2m and velocity estimation error is less than 0.5m/s.

\subsection{Joint Resource Allocation and Radar Receivers Selection}
In this section we mainly demonstrate the effects of radar receiver distribution and selection on the tradeoff between communication performance and detection performance. We set $M=2$, $N=1$, $K = 64$, $N_r = 4$, and the distance from the BS to the target is 300m.

In Fig.10 and Fig.11, we assume that $R_x$ radar receivers are evenly distributed on the circle with a radius of $radius$ centered around transmit BS. Fig.10 and Fig.11 shows the relationship between the transmission rate and the CRB of location with different $R_x$ and $radius$, respectively. In Fig.10, $radius$ is fixed to 50m and 4 receivers are chosen out of $R_x$ receivers to minimize CRB. The selective gain obtained by choosing optimal radar receiver set makes the transmit BS can allocate more power and bandwidth resource to communication users for given CRB constraints. Therefore, it is seen that increasing $R_x$ helps to improve transmission rate under same CRB constraints. However, for fixed $radius$, the performance gain will decreases with the increasing of the density of radar receivers. To reveal the relationship between the tradeoff performance and $radius$, Fig.11 exhibits the performance curves with varying $R_x$ and $radius$. The curves indicate that the performance improvement from increasing $radius$ is more significant than that from increasing $R_x$ due to increased spatial freedom and reduced detection distance.

In Fig.\ref{target_location1} and Fig.\ref{target_location2}, we assume that radar receivers are distributed around the target. In Fig.\ref{target_location1}, the radar receivers are distributed on a circle with a radius of 300m centered on the target. All radar receivers are uniformly distributed within a certain angle range which is denoted by the marker ``angle". For example, ``angle = 180, $R_x = 8$" means that 8 radar receivers are located on the circle with angle interval $180^o/8$. The larger the ``angle" is, the more dispersed the distribution of the receivers is. From Fig.\ref{target_location1}, we can see that compared with increasing $R_x$, increasing ``angle" (scattering receivers) can provide greater performance improvement. In Fig.\ref{target_location2}, we change the placement of receivers by locating half of them on a circle of 250m radius and the other half on a circle of 350m radius. In this case, the selected receivers may be on different circles (`non-equal distance distribution'). The curves with ``angle=360, $R_x=8$" and ``angle=360, $R_x=6$" in Fig.12 are also plotted for comparison. Fig.\ref{target_location2} shows that non-equal distance distribution of radar receivers results in a small performance gain compared to equal distance distribution. From Fig.\ref{target_location1} and Fig.\ref{target_location2} we can find that the minimum value of CRB is 0.05, which is much smaller than the minimum value of CRB (0.4) in Fig.10 and Fig.11. This is because when the radar receivers are distributed around the transmit BS, the minimum distance between the radar receivers is small. On the other hand, when the radar receivers are distributed around the target, especially when the radar receivers occupies the full angular range, the minimum distance between the radar receivers is maximized. Based on the above analysis, we can conclude: the increase of the number of available radar receivers and the increase of the relative distance between radar receivers can improve the estimation performance, and the later brings more estimation performance gain.

\begin{figure}[h]
\centering
\includegraphics[width=8cm] {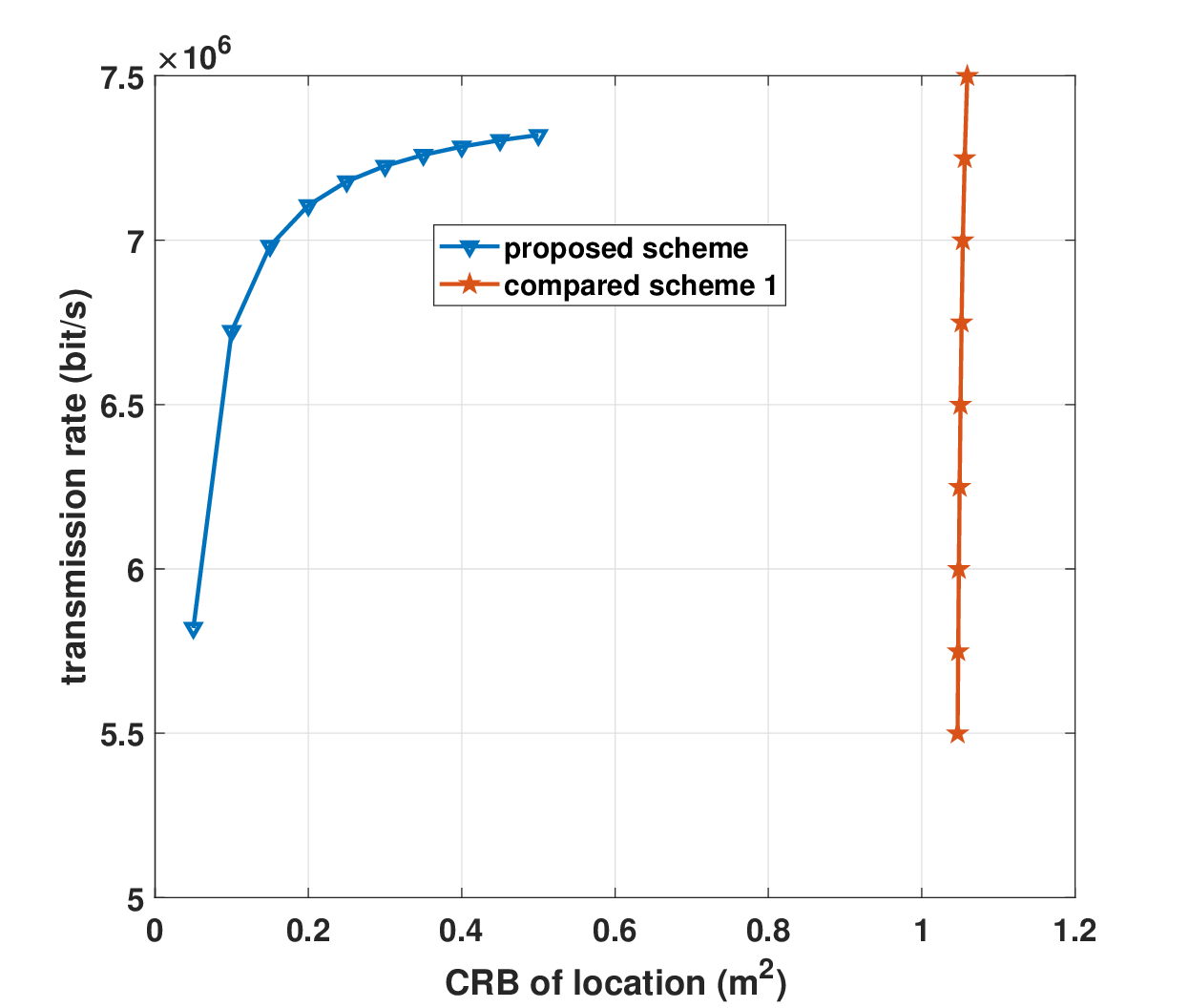}
\caption{Comparison of tradeoff performance with existing scheme.}
\label{compare1}
\end{figure}

\subsection{Comparison schemes}
In this subsection, we compare the proposed method with that in \cite{9124713}, which designed a joint transmit beamforming scheme for an integrated MIMO communication-radar system. Different from our frequency division scheme, the comparison scheme designed beamforming matrices for radar and communication waveforms on the sharing spectrum, respectively. Since \cite{9124713} did not consider OFDM systems, we extend the schemes in \cite{9124713} to OFDM systems in order to make a fair comparison with our scheme. In the comparison scheme, each subcarrier is used for both communication and detection, and the baseband OFDM signal on the $k$-th subcarrier is expressed as $\boldsymbol{s}_k = \boldsymbol{\Omega}_k^{R}\boldsymbol{b}_k^{R}+\boldsymbol{W}_{k}^C \boldsymbol b_{k}^C$ where $T_x \times T_x$ matrix $\boldsymbol{\Omega}_k^{R}$ is the beamforming matrix for radar waveform, the $T_x \times M$ matrix $\boldsymbol{W}_{k}^C$ is the communication precoder, $\boldsymbol{b}_k^{R}$ is radar signal and $\boldsymbol b_{k}^C$ is communication signal. We assume that the total power is equally allocated on each subcarrier. For each subcarrier, we design $\boldsymbol W_k^R$ and $\boldsymbol {\Omega}_k^R$ to optimize the beampattern under the constraint of signal-to-interference-plus-noise ratio (SINR) of communication signals with the method in \cite{9124713}. Then we utilize the optimized transmission parameters to calculated CRB (with method in sec.3.2) and communication rate, and the results are shown in Fig.14.

In Fig.14, we set $M=2$, $N=1$, $K=64$, $R_x=4$ and $N_r=4$. In our scheme, radar receivers are located on a circle with a radius of 50m centered on the transmit BS. The comparison scheme considered a monostatic radar system with $N_r$ receiver antennas. Locations of communication users and detection target are the same as that in Sec.5.1. In the comparison scheme, decreasing the SINR threshold of each communication user will improve the beam pattern used for detection of transmit signals. However, we find that the CRB decreases slightly when the designed detection beam pattern is improved. the gap between the actual and the ideal beampattern is reduced,  When the transmission rate is reduced to $5.5\text{Mbps}$, the CRB of location is about $1$, which is still larger than the CRB in our scheme. Therefore, with the same transmission rate, the proposed scheme is obviously superior to comparison schemes in terms of estimation performance. On the other hand, since the comparison scheme adopted space division multiplexing on each subcarrier, by optimizing beamforming vectors to control multiuser interference, it has the potential to achieve a higher transmission rate.

\section{Conclusion}
In this paper, we propose a distributed MIMO DFRC system which uses OFDM signals as integrated signals. Different subcarrier sets are used for communication and detection, respectively. To better achieve the detection function, we design the beamforming vector and sensing signals on sensing subcarriers and analyze the CRBs of target locations and velocities in different detection regions. Then, we maximize the transmission rate by jointly optimizing the power/subcarrier allocation scheme and radar receiver selection under CRB constraints. We propose an alternative optimization algorithm to solve the proposed MINLP problem, which solves the original problem by iteratively solving two subproblems. The first subproblem is related to power and subcarrier allocation, which is solved by combining SDP and DC approximation. The second subproblem is a quadratic integer problem related to radar receiver selection, which is equivalently transform into a convex quadratic integer problem that can be solved at a low complexity. Simulation results show the best tradeoff relationship between radar and communication performance by optimizing resource allocation and receivers selection.

\section*{ACKNOWLEDGEMENT}

This work was supported by the National Natural Science Foundation of China under grants 62072229, 62071220 and joint project of China Mobile Research Institute and X-NET.


\onecolumn
\appendix
\vspace{-9pt}
\subsection{Proof of Lemma 1}
\vspace{-9pt}
We will begin with equ.(10). When $\tau_{n,r}> T_{cp}$, the calculation of equ.(10) will involve two adjacent OFDM symbols. Therefore,
\begin{equation}
       \begin{aligned}
           \bar y_{r}(k,l)
		& = {1\over T} \int_{0} ^{T} y_r\left(t+\left( l-1 \right) T_{s} \right)e^{-j2\pi (k-1)\Delta ft}dt \\
		& = {1\over T} \int_{0} ^{\tau_{n,r}-T_{cp}}y_{r,l-1}\left(t +\left( l-1 \right) T_{s}\right)e^{-j2\pi (k-1)\Delta ft}dt +{1\over T} \int_{\tau_{n,r}-T_{cp}} ^{T} y_{r,l}\left(t +\left( l-1 \right) T_{s}\right)e^{-j2\pi (k-1)\Delta ft}dt,
       \end{aligned}
\end{equation}
where $y_{r,l}(t)=\sum_{n=1}^{N}\sum_{k=1}^{K}s_r c_{n,r}e^{j2\pi f_{n,r}t} e^{j2\pi (k-1)\Delta f (t-\tau_{n,r}-(l-1)T_s)}u(t-\tau_{n,r}-(l-1)T_s) \boldsymbol{\beta}_{k,n}^H\\ \times \sum_{n'=1}^N\sqrt{p_k}\sigma_{k,n'}^R\boldsymbol{\Omega}_{k,n'}^R\boldsymbol{b}_{k,n',l}^R + w(t)$. Substituting $y_{r,l}(t)$ into (A1), we have
\begin{equation}
       \begin{aligned}
                 &{1\over T} \int_{0} ^{\tau_{n,r}-T_{cp}}y_{r,l-1}\left(t+\left( l-1 \right) T_{s} \right)e^{-j2\pi (k-1)\Delta ft}dt\\
                =&{1\over T} \int_{0} ^{\tau_{n,r}-T_{cp}}\sum_{n=1}^{N}\sum_{k'=1}^{K}s_r c_{n,r} e^{j2\pi f_{n,r}\left(t+\left(l-1\right)T_s\right)} e^{j2\pi (k'-1)\Delta f (t-\tau_{n,r}+T_s)}\boldsymbol\beta_{k',n}^H\sum_{n'=1}^N\sqrt{p_{k'}}\sigma_{k',n'}^R\boldsymbol{\Omega}_{k',n'}^R\boldsymbol{b}_{k',n',l-1}^Re^{-j2\pi (k-1)\Delta f t}dt\\&+\bar w(k,l)\\
               \approx&{1\over T}\sum_{n=1}^{N}\sum_{k'=1}^{K}s_r c_{n,r}e^{j2\pi f_{n,r}\left(l-1\right)T_s}e^{-j2\pi (k'-1)\Delta f (\tau_{n,r}-T_s)}\boldsymbol\beta_{k',n}^H\sum_{n'=1}^N\sqrt{p_{k'}}\sigma_{k',n'}^R\boldsymbol{\Omega}_{k',n'}^R\boldsymbol{b}_{k',n',l-1}\int_{0} ^{\tau_{n,r}-T_{cp}}e^{j2\pi \left(k'-k\right)\Delta f t}dt\\&+\bar w(k,l),
       \end{aligned}\label{A1}
\end{equation}
and
\begin{equation}
       \begin{aligned}
                 &{1\over T} \int_{\tau_{n,r}-T_{cp}} ^{T}y_{r,l}\left(t+\left( l-1 \right) T_{s}  \right)e^{-j2\pi (k-1)\Delta ft}dt\\
               =&{1\over T} \int_{\tau_{n,r}-T_{cp}} ^{T}\sum_{n=1}^{N}\sum_{k'=1}^{K}s_rc_{n,r}e^{j2\pi f_{n,r}(t+(l-1)T_s)} e^{j2\pi (k'-1)\Delta f (t-\tau_{n,r})}\boldsymbol\beta_{k',n}^H\sum_{n'=1}^N\sqrt{p_{k'}}\sigma_{k',n'}^R\boldsymbol{\Omega}_{k',n'}^R\boldsymbol{b}_{k',n',l}e^{-j2\pi (k-1)\Delta ft}dt\\&+\bar w(k,l)\\
               \approx&{1\over T}\sum_{n=1}^{N}\sum_{k'=1}^{K}s_r c_{n,r}e^{j2\pi f_{n,r}(l-1)T_s}e^{-j2\pi (k'-1)\Delta f \tau_{n,r}}\boldsymbol\beta_{k',n}^H\sum_{n'=1}^N\sqrt{p_{k'}}\sigma_{k',n'}^R\boldsymbol{\Omega}_{k',n'}^R\boldsymbol{b}_{k',n',l}\int_{\tau_{n,r}-T_{cp}} ^{T}e^{j2\pi \left(k'-k\right)\Delta f t}dt+\bar w(k,l)\\
               =&{1\over T}\sum_{n=1}^{N}\sum_{k'=1}^{K}s_r c_{n,r}e^{j2\pi f_{n,r}(l-1)T_s}e^{-j2\pi (k'-1)\Delta f \tau_{n,r}}\boldsymbol\beta_{k',n}^H\sum_{n'=1}^N\sqrt{p_{k'}}\sigma_{k',n'}^R\boldsymbol{\Omega}_{k',n'}^R\boldsymbol{b}_{k',n',l}\times \\ &\left(\int_{0} ^{T}e^{j2\pi \left(k'-k\right)\Delta f t}dt-\int_{0} ^{\tau_{n,r}-T_{cp}}e^{j2\pi \left(k'-k\right)\Delta f t}dt\right)+\bar w(k,l)\\
               =&\sum_{n=1}^{N}s_r c_{n,r}e^{j2\pi f_{n,r}(l-1)T_s}e^{-j2\pi (k-1)\Delta f \tau_{n,r}}\boldsymbol\beta_{k,n}^H\sum_{n'=1}^N\sqrt{p_{k}}\sigma_{k,n'}^R\boldsymbol{\Omega}_{k,n'}^R\boldsymbol{b}_{k,n',l}-\\&{1\over T}\sum_{n=1}^{N}\sum_{k'=1}^{K}s_r c_{n,r}e^{j2\pi f_{n,r}(l-1)T_s}e^{-j2\pi (k'-1)\Delta f \tau_{n,r}}\boldsymbol\beta_{k',n}^H\sum_{n'=1}^N\sqrt{p_{k'}}\sigma_{k',n'}^R\boldsymbol{\Omega}_{k',n'}^R\boldsymbol{b}_{k',n',l}\int_{0} ^{\tau_{n,r}-T_{cp}}e^{j2\pi \left(k'-k\right)\Delta f t}dt+\bar w(k,l).
       \end{aligned}\label{A2}
\end{equation}
where approximations in (A2) and (A3) are because we approximate the phase rotation due to the doppler frequency within one OFDM block as constant \cite{100}. Based on (A2) and (A3), we have
\begin{equation}
       \begin{aligned}
           &\bar y_{r}(k,l)=\sum_{n=1}^{N}s_rc_{n,r}e^{j2\pi f_{n,r}(l-1)T_s}e^{-j2\pi (k-1)\Delta f \tau_{n,r}}\boldsymbol\beta_{k,n}^H\sum_{n'=1}^N\sqrt{p_k}\sigma_{k,n'}^R\boldsymbol{\Omega}_{k,n'}^R\boldsymbol{b}_{k,n',l}+{1\over T}\sum_{n=1}^{N}\sum_{k'=1}^{K}s_rc_{n,r}\times \\
           &\underbrace{e^{j2\pi f_{n,r}(l-1)T_s}e^{-j2\pi (k'-1)\Delta f \tau_{n,r}}\boldsymbol\beta_{k',n}^H\sum_{n'=1}^N\sqrt{p_{k'}}\sigma_{k',n'}^R\boldsymbol{\Omega}_{k',n'}^R\left(\boldsymbol{b}_{k',n',l-1}e^{j2\pi (k'-1)\Delta fT_s}-\boldsymbol{b}_{k',n',l}\right) \int_{0} ^{\tau_{n,r}-T_{cp}}e^{j2\pi \left(k'-k\right)\Delta f t}dt}_{\mathbf{ICI}}\\&+\bar w(k,l)
       \end{aligned}
\end{equation}
Obviously, when $\boldsymbol{b}_{k, n,l-1}e^{j2\pi (k-1)\Delta fT_s}=\boldsymbol{b}_{k, n,l}$, the ICI term is equal to $0$.

\subsection{The Expressions of $\mathbf{D}_{n'}(r,k,n)$ and $\mathbf{V}_{n'}(r,k,n)$}
\begin{equation}
	\begin{aligned}
    {[}\boldsymbol{D}_{n}{]}_{1,1} = \frac{8\pi^2\vert c_{n,r}\vert^2}{\sigma_{\bar{w}}^2}\boldsymbol{\beta}_{k,n}^H \boldsymbol{R}^*_{k,n'}\boldsymbol{\beta}_{k,n}\sum_{l=1}^{L}\Big(k^2\Delta f^2(\frac{\partial \tau_{n,r}}{\partial d_{n}^x})^2 - 2kl\Delta f T\frac{\partial f_{n,r}}{\partial d_{n}^x}\frac{\partial \tau_{n,r}}{\partial d_{n}^x}  + l^2 T^2 (\frac{\partial f_{n,r}}{\partial d_{n}^x})^2  \Big)\qquad \quad \label{D11}
	\end{aligned}
\end{equation}
\begin{equation}
	\begin{aligned}
{[}\boldsymbol{D}_{n}{]}_{1,2}  = \frac{8\pi^2\vert c_{n,r}\vert^2}{\sigma_{\bar{w}}^2}\boldsymbol{\beta}_{k,n}^H \boldsymbol{R}^*_{k,n'}\boldsymbol{\beta}_{k,n}\sum_{l=1}^{L}\Big(k^2\Delta f^2\frac{\partial \tau_{n,r}}{\partial d_{n}^x} \frac{\partial \tau_{n,r}}{\partial d_{n}^y} - kl\Delta f T(\frac{\partial f_{n,r}}{\partial d_{n}^x}\frac{\partial \tau_{n,r}}{\partial d_{n}^y}+\frac{\partial f_{n,r}}{\partial d_{n}^y}\frac{\partial \tau_{n,r}}{\partial d_{n}^x})+ l^2 T^2 \frac{\partial f_{n,r}}{\partial d_{n}^x}\frac{\partial f_{n,r}}{\partial d_{n}^y} \Big) \label{D12}
	\end{aligned}
\end{equation}
\begin{equation}
	\begin{aligned}
{[}\boldsymbol{D}_{n}{]}_{2,1}  = \frac{8\pi^2\vert c_{n,r}\vert^2}{\sigma_{\bar{w}}^2}\boldsymbol{\beta}_{k,n}^H \boldsymbol{R}^*_{k,n'}\boldsymbol{\beta}_{k,n}\sum_{l=1}^{L}\Big(k^2\Delta f^2\frac{\partial \tau_{n,r}}{\partial d_{n}^x} \frac{\partial \tau_{n,r}}{\partial d_{n}^y}
 - kl\Delta f T(\frac{\partial f_{n,r}}{\partial d_{n}^y}\frac{\partial \tau_{n,r}}{\partial d_{n}^x}+\frac{\partial f_{n,r}}{\partial d_{n}^x}\frac{\partial \tau_{n,r}}{\partial d_{n}^y}) + l^2 T^2 \frac{\partial f_{n,r}}{\partial d_{n}^x}\frac{\partial f_{n,r}}{\partial d_{n}^y} \Big)    \label{D21}
	\end{aligned}
\end{equation}
\begin{equation}
	\begin{aligned}
 {[}\boldsymbol{D}_{n}{]}_{2,2} = \frac{8\pi^2\vert c_{n,r}\vert^2}{\sigma_{\bar{w}}^2}\boldsymbol{\beta}_{k,n}^H \boldsymbol{R}^*_{k,n'}\boldsymbol{\beta}_{k,n}
\sum_{l=1}^{L}\Big(k^2\Delta f^2(\frac{\partial \tau_{n,r}}{\partial d_{n}^y})^2
- 2kl\Delta f T\frac{\partial f_{n,r}}{\partial d_{n}^y}\frac{\partial \tau_{n,r}}{\partial d_{n}^y}  + l^2 T^2 (\frac{\partial f_{n,r}}{\partial d_{n}^y})^2 \Big)\qquad\quad \label{D22}
	\end{aligned}
\end{equation}
\begin{equation}
	\begin{aligned}
     & [\boldsymbol{V}_{n}]_{1,1} = \frac{8\pi^2\vert c_{n,r}\vert^2}{\sigma_{\bar{w}}^2}\boldsymbol{\beta}_{k,n}^H \boldsymbol{R}^*_{k,n'}\boldsymbol{\beta}_{k,n}\sum_{l=1}^L l^2 T^2(\frac{\partial f_{n,r}}{\partial v_{n}^x})^2  \\ \label{V11}
	\end{aligned}
\end{equation}
\begin{equation}
	\begin{aligned}
     & [\boldsymbol{V}_{n}]_{1,2} = \frac{8\pi^2\vert c_{n,r}\vert^2}{\sigma_{\bar{w}}^2}\boldsymbol{\beta}_{k,n}^H \boldsymbol{R}^*_{k,n'}\boldsymbol{\beta}_{k,n}\sum_{l=1}^L l^2 T^2 \frac{\partial f_{n,r}}{\partial v_{n}^x}   \frac{\partial f_{n,r}}{\partial v_{n}^y} \\ \label{V12}
	\end{aligned}
\end{equation}
\begin{equation}
	\begin{aligned}
     & [\boldsymbol{V}_{n}]_{2,1} = \frac{8\pi^2\vert c_{n,r}\vert^2}{\sigma_{\bar{w}}^2}\boldsymbol{\beta}_{k,n}^H \boldsymbol{R}^*_{k,n'}\boldsymbol{\beta}_{k,n}\sum_{l=1}^L l^2 T^2 \frac{\partial f_{n,r}}{\partial v_{n}^x}  \frac{\partial f_{n,r}}{\partial v_{n}^y}  \\ \label{V21}
	\end{aligned}
\end{equation}
\begin{equation}
	\begin{aligned}
     & [\boldsymbol{V}_{n}]_{2,2} = \frac{8\pi^2\vert c_{n,r}\vert^2}{\sigma_{\bar{w}}^2}\boldsymbol{\beta}_{k,n}^H \boldsymbol{R}^*_{k,n'}\boldsymbol{\beta}_{k,n}\sum_{l=1}^L l^2 T^2(\frac{\partial f_{n,r}}{\partial v_{n}^y})^2  \\ \label{V22}
	\end{aligned}
\end{equation}

\subsection{Proof of Lemma 2}

\begin{equation}
	\begin{aligned}
	\quad \Big[(\sum_{i}a_i\boldsymbol B_i)\Big]_{1,1}^{-1}=\frac{\Big[\sum_{i}a_i\boldsymbol B_i\Big]_{2,2}}{det(\sum_{i}a_i\boldsymbol B_i)}
&=\frac{\sum_{i}a_i [\boldsymbol B_i]_{2,2}}{\sum_{i}a_i\left[\boldsymbol B_i\right]_{1,1}\sum_{i}a_i\left[\boldsymbol B_i\right]_{2,2}-\sum_{i}a_i\left[\boldsymbol B_i\right]_{1,2}\sum_{i}a_i\left[\boldsymbol B_i\right]_{2,1}}\\
&=\frac{\sum_{i}a_i [\boldsymbol B_i]_{2,2}}{\sum_{i}\sum_{i'}a_i a_{i'}\left(\left[\boldsymbol B_i\right]_{1,1}\left[\boldsymbol B_{i'}\right]_{2,2}-\left[\boldsymbol B_i\right]_{1,2}\left[\boldsymbol B_{i'}\right]_{2,1}\right)}\\
&=\frac{\boldsymbol {a}^{T}\boldsymbol{p}_1}{ \boldsymbol{a}^{T} \boldsymbol{Q} \boldsymbol{a}}
	\end{aligned}
\end{equation}
where $\boldsymbol{p}_1$ and $\boldsymbol{Q}$ are defined in Lemma 2.

\twocolumn

\bibliographystyle{gbt7714-numerical}
\bibliography{myref}




\end{document}